\documentclass[twocolumn]{article}

\usepackage{amsmath}
\usepackage{amssymb}

\usepackage{bm}
\usepackage[T1]{fontenc}
\usepackage[left=2.00cm, right=2.00cm, top=1.50cm, bottom=2.00cm]{geometry}
\usepackage{graphicx}
\usepackage{latexsym}
\usepackage{mathrsfs, mathtools}
\usepackage{multirow}
\usepackage[utf8]{inputenc}
\usepackage{natbib}
\usepackage{subfig}
\usepackage{textcomp}
\usepackage{url}
\usepackage{titling}

\let\notORI\not % Needed for \neq
\usepackage[mathb]{mathabx}
\let\not\notORI

\bibpunct{(}{)}{;}{a}{}{,}    %to follow the A&A style 
\usepackage[varg]{txfonts}
\graphicspath{{.}}

\usepackage{authblk}

\usepackage{tikz,pgfplots}
\usetikzlibrary{arrows,intersections,backgrounds}
\usetikzlibrary{positioning,calc,plotmarks}

\newsavebox\affbox

\graphicspath{{.}}

\def\R{\mathbb{R}}
\title{\textbf{Short arc orbit determination and imminent impactors in
    the Gaia era}}

\author[1,2]{F.~Spoto}
\author[3,4]{A.~Del Vigna}
\author[3]{A.~Milani}
\author[3]{G.~Tommei}
\author[1]{P.~Tanga}
\author[1]{F.~Mignard}
\author[1]{B.~Carry}
\author[2]{W.~Thuillot}
\author[2]{D.~Pedro}

\affil[1]{Université C\^ote d'Azur, Observatoire de la C\^ote d'Azur,
  CNRS, Laboratoire Lagrange, route de l'Observatoire, Nice,
  France}
\affil[2]{IMCCE, Observatoire de Paris, PSL Research University,
  CNRS, Sorbonne Universités, UPMC Univ. Paris 06, Univ. Lille, 77
  av. Denfert-Rochereau F-75014 Paris, France}
\affil[3]{Dipartimento di Matematica, Universit\`a di Pisa, Largo
  Bruno Pontecorvo 5, Pisa, Italy}
\affil[4]{Space Dynamics Services s.r.l, via Mario Giuntini, Navacchio di Cascina, Pisa, Italy}

\date{\today}

\begin{document}
\maketitle

{Short-arc orbit determination is crucial when an asteroid is first
  discovered. In these cases usually the observations are so few that
  the differential correction procedure may not converge. We have
  developed an initial orbit computation method, based on the
  systematic ranging, an orbit determination techniques which
  systematically explores a raster in the topocentric range and
  range-rate space region inside the admissible region. We obtain a
  fully rigorous computation of the probability for the asteroid that
  could impact the Earth within few days from the discovery, without
  any a priori assumption. We test our method on the two past
  impactors $2008~TC_3$ and $2014~AA$, on some very well known cases,
  and on two particular objects observed by the ESA Gaia mission.}


%------------------------%
%      INTRODUCTION      %
%------------------------%
\section{Introduction}
\label{sec:intro}
Short-arc orbit determination is a very important step when an
asteroid is first discovered. In these cases the timing is essential,
because we are interested in a rapid follow-up of a possible imminent
impactor, which is an asteroid impacting the Earth shortly after its
discovery, within the same apparition (interval of observability). The
observations are so few that the standard differential
correction procedure \citep{milani:orbdet} to find an orbit by a
least-squares minimization fails, and other methods need to
be used to extract information on the orbit of the object.

Several initial orbit computation methods have been developed in the
last 25 years. For instance, \cite{muinonen93} define a Gaussian
probability density on the orbital elements space using the Bayesian
inversion theory. In particular, they determine asteroid orbital
elements from optical astrometric observations using both a
  priori and a posteriori densities, the latter computed with
a Monte Carlo method.

The few observations in the short arc constrain the position of the
object in the sky, but they leave almost unknown the distance from the
observer (topocentric range) and the radial velocity (topocentric
range-rate). Thus, ranging methods have been developed over
the years to replace or refine the Monte Carlo approach in
the short arc orbit determination. There are two alternative
approaches to the ranging methods: the statistical and the systematic
ones.

The original statistical ranging method (\cite{virtanen2001},
\cite{muinonen2001}) starts from the selection of a pair of
astrometric observations. Then, the the topocentric ranges at
  the epoch of the observations are randomly sampled. Candidate
orbital elements are included in the sample of accepted elements if
the $\chi^2$ value between the observed and computed observations is
within a pre-defined threshold. \cite{oszkiewicz2009} improve the
statistical ranging using the Markov-Chain Monte Carlo (MCMC)
  to sample the phase space. The MCMC orbital ranging method is based
on a bi-variate Gaussian proposal PDF for the topocentric ranges. Then,
\cite{muinonen2016} develop a random-walk ranging method in
which the orbital-element space is uniformly sampled, up to a $\chi^2$
value, with the use of the MCMC method. The weights of each set of
orbital elements are based on a posteriori probability density value
and the MCMC rejection rate. They have developed this method for the
ESA Gaia mission, in the framework of Gaia alerts on potentially new
discovered objects by Gaia (see \cite{tanga2016}).

On the other side, \cite{chesley2005} and \cite{farnocchia2015}
introduce the so-called systematic ranging, which systematically
explores a raster in the topocentric range and range-rate space
$(\rho, \dot{\rho})$. This technique permits to describe the asteroid
orbital elements as a function of range and range-rate. Then, the
systematic ranging also allows one to determine the subset of the
sampling orbits which leads to an impact with the Earth.

In this paper we describe a new approach to the systematic ranging,
based on the knowledge of the Admissible Region (AR)
\citep{milani2004AR}, and a new method to scan the region. The
process has some main advantages if compared to the other methods
described above.

\begin{enumerate}
\item Our grid is more efficient, for two main reasons:
  \begin{itemize}
  \item[-] we discard all the objects that are not in the AR, saving
    CPU time and making the systematic ranging more accurate
      in finding the region in the $(\rho, \dot{\rho})$ space of the
      possible orbital solutions;
  \item[-] we use two different grids depending on the boundary of the
    AR. The first grid is larger and less dense, the second one is
    based on a refinement using the value of the post-fit $\chi^2$ of
    each point in the first grid (see Section~\ref{sec:adm_reg}).
  \end{itemize}
\item The computation of the probability for the potential impactors
  is a rigorous probability propagation from the astrometric error
  model, without any assumption of a priori probability density
  function on the range/range-rate space (see Section~\ref{sec:PDF}).
\end{enumerate}

In Section~\ref{sec:adm_reg} we summarize the features of the AR, and
describe the sampling of the $(\rho, \dot{\rho})$ space needed for the
creation of the Manifold Of Variations (MOV), an extension of the Line
Of Variations (LOV, \cite{milani2005multsol,milani:orbdet}). In
Section~\ref{sec:cobweb} we present the so-called spider-web
\citep{tommei:phd}, to be used when a nominal orbit is
available. Section~\ref{sec:PDF} describes how to compute the impact
probability when an impactor is found. In Section~\ref{sec:results} we
apply our method to the well known cases $2008~TC_3$ and $2014~AA$,
and we show other examples to explore the capabilities of the new
method. Then in Section~\ref{sec:gaia} we test our method on new
objects discovered by Gaia in the framework of the Gaia alerts, and we
show the points of strength of this approach applied to Gaia
observations. Section~\ref{sec:conclusions} contains our conclusions
and comments.

%---------------------------%
%     ADMISSIBLE REGION     %
%---------------------------%
\section{Sampling of the topocentric range and range-rate space}
\label{sec:sampling}

\subsection{Admissible Region and systematic ranging}
\label{sec:adm_reg}
Even if the observations are too scarce, we can anyway
compute the right ascension $\alpha$, the declination $\delta$, and
their time derivatives $\dot{\alpha}$ and $\dot{\delta}$, by fitting
both angular coordinates as a function of time with a polynomial
model. This four quantities can be assembled together to form the
attributable \citep{milani2005}:
\begin{equation}\label{eq:attributable}
  \mathcal{A} = (\alpha, \delta, \dot{\alpha}, \dot{\delta}) \in
  \mathcal{S}^1 \times (-\pi/2, \pi/2) \times \R^2
\end{equation}
at a chosen time $\bar t$, which could be the time of the first
observation or the mean of the observation times. The information
contained in the attributable leaves completely unknown the
topocentric distance $\rho$ and the radial velocity $\dot{\rho}$. We
would have a full description of the topocentric position and velocity
of the asteroid in the attributable elements $(\alpha, \delta,
\dot{\alpha}, \dot{\delta}, \rho, \dot{\rho})$, if $\rho$ and
$\dot{\rho}$ were known. From now on, we will use
$\mathbf{x}=(\mathcal A, \bm{\rho})$, where
$\bm{\rho}=(\rho,\dot\rho)$, to describe the attributable elements.

Given an attributable, we can define the AR as the set of all the
possible couples $(\rho, \dot{\rho})$ satisfying the following
conditions (see \cite{milani2004AR} for more mathematical details):
\begin{enumerate}
  \item\label{cond:solsis} the object belongs to the Solar System, and
    it is not a too long period comet. We consider only the objects
    for which the value of the heliocentric energy is less than
    $-k^2/(2a_{max})$, where $a_{max}=100$~au and $k=0.01720209895$ is
    the Gauss' constant;
  \item\label{cond:earth} the corresponding object is not a satellite
    of the Earth, i.e. the orbit of the object has a non-negative
    geocentric energy when inside the sphere of influence of
      the Earth, whose radius is
\[
    R_{SI} = a_\Earth \sqrt[3]{\frac{\mu_\Earth}{3\mu_\Sun}} \simeq 0.010044\text{
      au}.
\]

\end{enumerate}
The AR is a compact set, and it can have at most two connected
components, which means that it could be represented as the union of
no more than two disjoint regions in the $(\rho,\dot{\rho})$
space. The AR has usually one component, and the case with two
components indicates the possibility for the object to be distant
(perihelion $q>28$ au). As shown in ~\cite{milani2004AR}, the number
of connected components depends on the number of the roots of a
polynomial resulting from condition number~\ref{cond:solsis}. This
polynomial cannot have more than three distinct real positive roots:
the AR has two connected components if the roots are
three, and it has one component if there is only one root. It is worth
noting that the region defined by condition number~\ref{cond:solsis}
could contain points with arbitrarily small values of $\rho$. The
boundary of the region given by condition~\ref{cond:earth} turns out
to have two different shapes: it can be formed just by the line of
geocentric energy equal to $0$ (if it is entirely contained in the
region $0<\rho<R_{SI}$), or by a segment of the straight vertical line
$\rho=R_{SI}$ and two arcs of the zero-curve of the geocentric energy
(for $0<\rho<R_{SI}$). We also discard the orbits corresponding to
meteors too small to be source of meteorites, using the condition
$H\leq H_{max}$, where $H_{max} = 34.5$ is the \textit{shooting star
  limit}~\citep{milani2004AR}, and $H$ is the absolute
magnitude. Given all these conditions, we can sample the AR with a
finite number of points.

If a nominal solution does not exist, we make use
of the systematic ranging to scan the AR. We sample the
AR in two different ways, depending on the number of connected
components, and the values of the roots ($r_1$, $r_2$ and $r_3$, in
ascending order). Table~\ref{tab:summary_AR} summarizes the conditions
and the grids used in the $(\rho,\dot{\rho})$ space. In particular, we
compute a rectangular grid in the range/range-rate space, with range
as in Table~\ref{tab:summary_AR}, and range-rate controlled by the AR
equations~\citep{milani2004AR}. Nevertheless, since the AR has a shape
dictated by a polynomial equation and it is not a rectangle, we check
the value of the heliocentric energy for each grid point, and we
discard those not satisfying condition~\ref{cond:solsis}. Orbits not
satisfying condition~\ref{cond:earth} are discarded as well, except
when we compute the probability for the asteroid to be a satellite of
the Earth\footnote{The object could be either an artificial satellite
  or an interplanetary orbit in a temporary Earth satellite
  capture~\citep{granvik2012}.}.

\begin{table}[h]
  \begin{center}
    \caption{The table shows different methods we use to sample
      the AR in the $(\rho,\dot{\rho})$ space with
      respect to the values of the roots, and the connected components
      of the AR. The sampling in $\dot\rho$ is always uniform.}
     \resizebox{\columnwidth}{!}{\begin{tabular}{l|c|c|l}
      \hline
      \multirow{2}*{\textbf{Roots}} & \textbf{AR} &  \multirow{2}*{\textbf{Grid}} & \multirow{2}*{\textbf{Sampling in} $\mathbf{\rho}$}\\
      & \textbf{components} &\\
      \hline
      $r_1 < \sqrt{10}$ au & 1 & $50\times 50$ & Unif. in $\log_{10}(\rho)$\\
      $r_1 \ge \sqrt{10}$ au & 1 & $50\times 50$ & Unif. in $\rho$\\
      $r_1 > 0$, $r_2 > 0$, $r_3 > 0$ & 2 & $100\times 100$ & Unif. in $\rho$\\
      \hline
    \end{tabular}}
    \label{tab:summary_AR}
  \end{center}
\end{table}

The target function is defined by
\[
    Q(\mathbf{x}) = \frac{1}{m} \bm{\xi}(\mathbf x)^T W \bm{\xi}(\mathbf x),
\]
where $\mathbf{x}=(\mathcal A,\bm\rho)$ are the fit parameters, $m$ is
the number of observations used in the least squares fit, $\bm{\xi}$
is the vector of the observed-computed debiased astrometric
residuals\footnote{In case there is a bias in the
  observations~\citep{farnocchia:fcct}, the residuals are computed
  following the classical definition of the residuals as
  observed-computed, and also subtracting the biases vector.}, and $W$
is the weight matrix. The choice of the weights for each observatory
is fundamental, and it has to take into account the debiasing of the
star catalog systematic errors, unless the astrometric reduction has
already been performed with an essentially bias-free star catalog,
e.g. the Gaia DR1~\citep{gaia_astrometry_2016}.

Given a subset $K$ of the AR, we define the Manifold Of
Variations $\mathcal M$ as the set of the points
$(\mathcal A^*(\bm\rho_0),\bm\rho_0)$ such that $\bm\rho_0\in K$ and
$\mathcal A^*(\bm\rho_0)$ is the local minimum of the function
$\left.Q\right|_{\bm\rho=\bm\rho_0}$. In addition, the value of the
minimum RMS of the residuals is less than a given threshold
$\Sigma$. In general, the MOV is a 2-dimensional manifold, such that
the differential of the map from the sampling space to $\mathcal M$
has rank $2$.

In the case of the systematic ranging, $K$ is the AR,
scanned with a regular semi-logarithmic or uniform grid. For each
sample point $\bm\rho_0=(\rho_0,\dot\rho_0)$ we fix $\rho=\rho_0$ and
$\dot\rho=\dot\rho_0$ in the target function, and then we search
$\mathcal A^*(\bm\rho_0)$ by means of an iterative procedure, the
doubly constrained differential corrections. The normal equation is
\[
    C_{\mathcal A} \Delta \mathcal A = D_{\mathcal A},
\]
where
\[
    C_{\mathcal A} = B_{\mathcal A}^T W B_{\mathcal A}\,,\quad
    D_{\mathcal A} = -B_{\mathcal A}^T W \bm{\xi}\,, \quad
    B_{\mathcal A} = \frac{\partial{\bm{\xi}}}{\partial{\mathcal A}}.
\]
We indicate as $K'$ the subset of $K$ on which the doubly constrained
differential corrections converges. In this way, the sampling of the
MOV is done over $K'$.

For each point $\mathbf{x}$ on the MOV, we also compute a $\chi$ value
\begin{equation}{\label{chimov}}
    \chi(\mathbf x) = \sqrt{m(Q(\mathbf x)- Q^*)},
\end{equation}
where $Q^*$ is the minimum value of the target function:
$Q(\mathbf{x}^*)$ if a reliable nominal solution exists, or the
minimum value of $Q(\mathbf{x})$ over $K'$ otherwise. 

When a nominal solution does not exist, the systematic ranging is
performed by a two-step procedure. The first iteration is to compute a
grid following the conditions in Table~\ref{tab:summary_AR}. Once we
obtain a first preliminary grid, we densify for a higher
resolution. We select the minimum and the maximum value of $\rho$ and
$\dot{\rho}$ among all the values of the points for which we have a
convergence of the $4$-dimension differential correction, and the
value of $\chi$ is less than $5$. We also compute the score with
respect to the first grid, and we use the value to select the grid for
the second step. The score gives us a first insight into the
  nature of the object, even when the asteroid is not a potential
  impactor. We define the score as a probability of the object to
  belong to different classes (NEO, MBO, DO, and SO), where each class
  is defined by the following conditions:

\begin{itemize}
\item \textbf{NEO}: Near Earth Object, an object with $q < 1.3$ au,
  where $q$ is the perihelion distance (in au).
\item \textbf{MBO}: Main Belt Object, belonging either to the Main
  Belt or to the Jupiter Trojans. In particular we choose the
  conditions
  \[
    \left\{\begin{array}{ll} 1.7\text{ au} < a < 4.5\text{ au}\\ e<0.4\end{array}\right.
    \quad\text{or}\quad
    \left\{\begin{array}{ll} 4.5\text{ au} < a < 5.5\text{ au}\\ e<0.3\end{array}\right.
  \]
  where $a$ is the semi-major axis (in au), and $e$ is the
  eccentricity.
\item \textbf{DO}: Distant Object, characterized by $q > 28$ au
  (e.g. a Kuiper Belt Object (KBO)).
\item \textbf{SO}: Scattered Object, not belonging to any of the
  previous classes.
\end{itemize}

If the object is close to be a NEO (the NEO-score is more than
$50\%$), we use a uniform grid in $\log_{10}(\rho)$, otherwise we use
a uniform grid in $\rho$. Then we compute again the Manifold Of
Variations on the new and denser $100 \times 100$ grid.

%--------------------%
%       COBWEB       %
%--------------------%
\subsection{Spider web}
\label{sec:cobweb}
Let us suppose that a nominal orbit has been obtained by unconstrained
differential corrections, starting from a preliminary orbit as first
guess (for instance using the Gauss' method,
~\citep{milani:orbdet}). Then, we can use the nominal solution as the
center of the subset of the MOV we are interested in, and we can adopt
a different procedure to scan the AR. If a nominal orbit exists, and
the value of the geodesic curvature signal-to-noise ratio
(SNR)~\citep{milani2008} is greater than $3$, instead of using a grid
(Section~\ref{sec:adm_reg}), we compute a spider web sampling in a
neighborhood of the nominal solution \citep{tommei:phd}. This is
obtained by following the level curves of the quadratic approximation
of the target function used to minimize the RMS of the observational
residuals. The advantage of the use of the cobweb is that, firstly, it
is faster than the systematic ranging, and secondly it is more
accurate in the cases for which we have already a reliable nominal
solution.

\begin{figure}[h]
  \centering
  \includegraphics[width=\columnwidth]{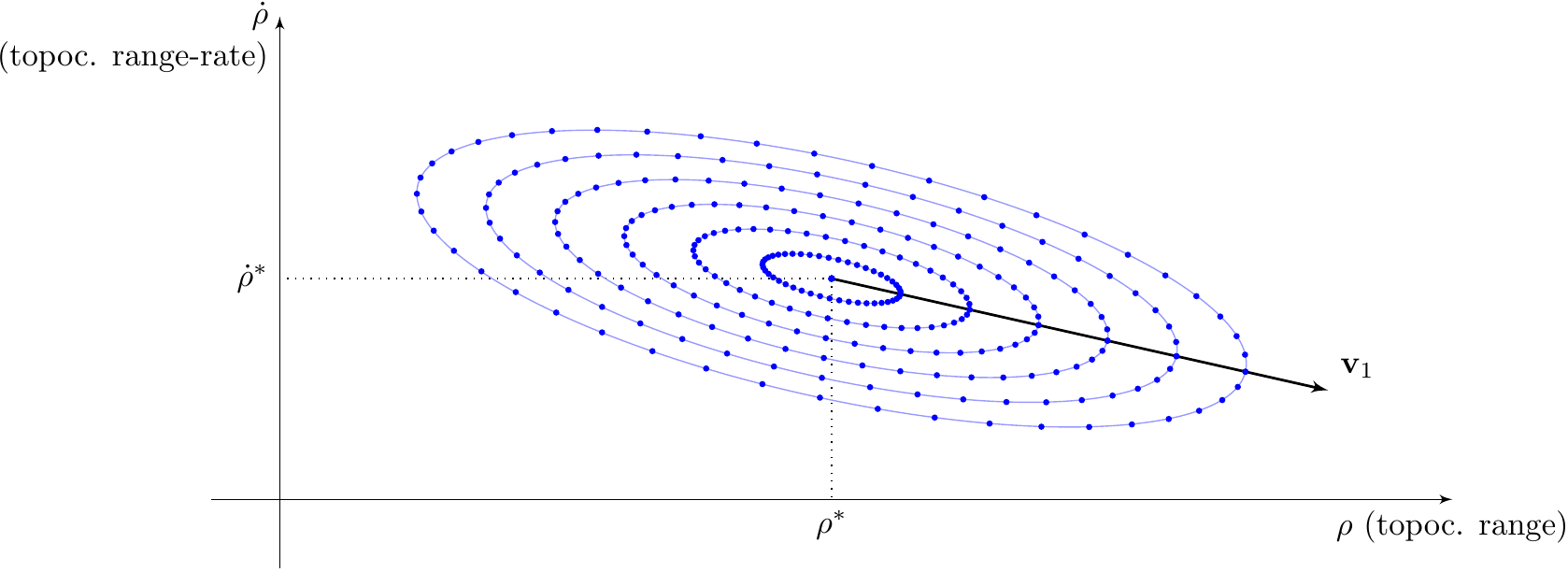}
  \caption{The figure shows an example of spider web around the
    nominal solution $\bm \rho^*=(\rho^*,\dot\rho^*)$. The points follow
    concentric ellipses, corresponding to different values of the
    parameter $R$. For each fixed direction (identified by $\theta$),
    there is one point on each level curve.}
  \label{fig:cobweb}
\end{figure}

Let $\mathbf{x}^*$ be the nominal solution with its uncertainty,
represented by the $6\times 6$ covariance matrix $\Gamma$. In a
neighborhood of $\mathbf{x}^*$, the target function can be well
approximated by means of the quadratic form defined by the normal
matrix $C=\Gamma^{-1}$. The matrix $C$ is positive definite, hence the
level curves of the target function are concentric 5-dimensional
ellipsoids in the 6-dimensional orbital elements space. The level
curves on the $(\rho,\dot\rho)$ space are represented by the marginal
ellipsoids, defined by the normal matrix
\[
C^{\bm\rho\bm\rho}=\Gamma_{\bm\rho\bm\rho}^{-1},
\]
where $\Gamma_{\bm\rho\bm\rho}$ is the restriction of $\Gamma$ to the
$(\rho,\dot\rho)$ space. To sample these curves we choose the maximum
value $\sigma_{max} = 5$ for the confidence parameter. Then, for each
level curve within the confidence level $\sigma_{max}$, we select the
points corresponding to some fixed directions. We initially create a
regular grid of points in the space of polar elliptic coordinates
$(R,\theta)$, where $0\leq \theta < 2\pi$ and $0\leq R \leq
\sigma_{max}$\footnote{We assume that the nominal solution
  corresponds to the point $(0,0)$.}. Then we apply to each point of
the grid the following transformation, depending on the covariance
matrix of the nominal orbit and on the orbit itself:
\begin{equation}\label{eq:cobweb}
  \begin{pmatrix}
    \rho \\
    \dot{\rho}
  \end{pmatrix}
  = R
  \begin{pmatrix}
    \sqrt{\lambda_1} \cos{\theta} & -\sqrt{\lambda_2} \sin{\theta}\\
    \sqrt{\lambda_2} \sin{\theta} & \phantom{-}\sqrt{\lambda_1} \cos{\theta}
  \end{pmatrix}
  \mathbf{v_1} +
  \begin{pmatrix}
    \rho^* \\
    \dot{\rho}^*
  \end{pmatrix},
\end{equation}
where $\lambda_1>\lambda_2$ are the eigenvalues of the $2 \times 2$
matrix $\Gamma_{\bm\rho\bm\rho}$, $\mathbf{v_1}$ is the eigenvector
corresponding to the greatest eigenvalue $\lambda_1$, and $\rho^*$ and
$\dot{\rho}^*$ are the range and range-rate values of the nominal
solution. Figure \ref{fig:cobweb} shows an example of spider web
sampling on the $(\rho,\dot\rho)$ plane.

%--------------------------------%
%       PROBABILITY DENSITY      %
%--------------------------------%
\section{Probability density computation}
\label{sec:PDF}
We obtain a probability distribution on the sampling space to be used
for several applications, such as the computation of the impact
probability or the score. We begin assuming that the residuals are a
Gaussian random variable $\bm\Xi$, with zero mean and covariance
$\Gamma_{\bm\xi}=W^{-1}$. Hence the probability density function on
the residuals space is
\begin{align}\label{eq:res_gauss}
    p_{\bm\Xi}(\bm \xi) &= N(\bm 0,\Gamma_{\bm \xi})(\bm \xi)=
    \frac{\sqrt{\det W}}{(2\pi)^{m/2}}
    \exp\left(-\frac{mQ(\bm\xi)}{2}\right)=\nonumber\\ &=\frac{\sqrt{\det
        W}}{(2\pi)^{m/2}} \exp\left(-\frac 12 \bm\xi^TW\bm\xi\right)
\end{align}

A possible approach to propagate the density \eqref{eq:res_gauss} to
the sampling space uses the Bayesian theory to combine the density
coming from the residuals with a prior distribution. The {\em a
  posteriori} probability density function for $(\rho, \dot{\rho})$ is
given in \cite{muinonen93} as
\begin{displaymath}
  p_{post}(\rho, \dot{\rho}) \propto p(\bm{\xi}(\rho, \dot{\rho}))
  \cdot p_{prior}(\rho, \dot{\rho})
\end{displaymath}
where $p_{prior}$ is a prior distribution on the sampled
space. Hereinafter we report some possible choices for the prior
probability.

\begin{itemize}
\item \emph{Jeffreys' prior}. It has been used for the first time in
  \cite{muinonen2001}. It takes into account the partial derivatives
  of the vector of the residuals with respect to the coordinates
  $(\rho, \dot{\rho})$. Jeffreys' prior tends to favor orbits where
  the object is close to the observer, because of the sensitivity of
  the residuals for small topocentric distances. \cite{granvik2009}
  makes versions of the code (which utilizes Jeffreys' prior) publicly
  available for the first time.
\item \emph{Prior based on a population model}. This approach requires
  the computation of the prior distribution as posterior of another
  prior, which is selected as proportional to $\rho^2$ by geometric
  considerations on the Cartesian space of position and velocity.
\item \emph{Uniform distribution}. Uniform distribution in the
  $(\rho,\dot\rho)$ space.
\end{itemize}
\cite{farnocchia2015} gives a detailed description of all these
different possible choices, and they also analyze how the impact
probabilities change according to different prior distributions. They
conclude that the uniform distribution is a good choice for an a
  priori probability density function, because it represents a good
compromise between a simple approach and the identification of
potential impactors.

Hereinafter we propose a new method to propagate the probability
density function $p_{\bm\Xi}(\bm\xi)$ back to the sampling space. This
method is a rigorous propagation of the density function according to
the probability theory, and it does not use any a priori
assumption. The main idea is that we propagate the probability density
function from the residual space to the orbital element space (more
precisely, on the Manifold of Variations), and then to the sampling
space, according to the Gaussian random variable transformation
law. The main difference in using this approach instead of the uniform
distribution, is that we compute the Jacobian determinant of the
transformation and it is not equal to 1, which is the value chosen by
\cite{farnocchia2015}.

We define the following spaces:
\begin{itemize}
\item $S$ is the space of the sampling variables. It changes depending
  on the case we are considering: $S=\R^+ \times \R$ if the sampling
  is uniform in $\rho$, $S=\R^2$ if the sampling is uniform in
  $\log_{10}\rho$, and $S=\R^+ \times \mathcal S^1$ in the cobweb
  case.
\item $K'$ is the subset of the points of the AR such
  that the doubly constrained differential corrections give a point on
  the MOV;
\item $\mathcal M$ is the MOV, a 2-dimensional manifold in the
  six-dimensional orbital elements space $\mathcal{X}$;
\item $\R^m$ is the residuals space. The residuals are a function of
  the fit parameters: $\bm \xi=F(\bold x)$, with
  $F:\mathcal{X}\rightarrow \R^m$ differentiable, and we define the
  manifold of possible residuals as $V=F(\mathcal{X})$ \citep[Section
  5.7]{milani:orbdet}.
\end{itemize}
Without loss of generality, we can assume that
\begin{equation}\label{eq:res_norm_gauss}
    p_{\bm\Xi}(\bm \xi) = N(\bm 0,I_m)(\bm \xi)= \frac
    1{(2\pi)^{m/2}}\exp\left(-\frac 12 \bm\xi^T\bm\xi\right),
\end{equation}
where $I_m$ is the $m\times m$ identity matrix. As explained in
\citet[Section 5.7]{milani:orbdet}, this is obtained by using the
normalized residuals in place of the true residuals; biases due to
star catalog can also be removed while forming the normalized
residuals. With this technique, the probability density function
becomes normalized to a standard normal distribution. Thus from now
on, we will use $\bm\xi$ to indicate the normalized residuals, and the
function $F$ maps the orbital elements space to the normalized
residuals space.

Then we consider the following chain of maps (defined in
\ref{app:IPcomp})
\[
S \overset{f_\sigma}{\longrightarrow} \R^+\times \R \supseteq K'
\overset{f_\mu}{\longrightarrow} \mathcal{X} \supseteq \mathcal M
\xrightarrow{\left.F\right|_{\mathcal M}}V
\]
and we use their Jacobian matrices to compute the probability density
function on $S$. Let $\mathbf{s}$ be the variable of the sampling
space $S$, and let $\mathbf{S}$ be the corresponding random
variable. We use the compact notation $\chi^2(\mathbf{s})$ to indicate
$\chi^2(\mathbf{x}(\bm\rho(\mathbf{s})))$. The probability density
function of $\mathbf{S}$ is
\tiny
\begin{equation}\label{eq:p_S}
p_{\mathbf{S}}(\mathbf{s}) = \frac{\displaystyle
  \exp\left(-\frac{\chi^2(\mathbf{s})}{2}\right) \det
  M_\mu(\bm\rho(\mathbf{s}))\det M_\sigma(\bm\rho(\mathbf{s}))}
{\displaystyle\int_{f_{\sigma}^{-1}(K')}
  \exp\left(-\frac{\chi^2(\mathbf{s})}{2}\right)\det
  M_\mu(\bm\rho(\mathbf{s}))\det
  M_\sigma(\bm\rho(\mathbf{s}))\,d\mathbf{s}}
\end{equation}
\normalsize
where $M_\mu$ is the $2\times 2$ matrix associated to $f_\mu$
(considered as tangent map to the surface $\mathcal M$), and
$M_\sigma$ is the $2\times 2$ Jacobian matrix of $f_\sigma$. The
derivation of \eqref{eq:p_S} is given in \ref{app:IPcomp} (explicit
expressions for the Jacobian determinants are provided in
\ref{subsec:MOV-AR} and \ref{subsec:AR-S}, respectively).

It is worth noting that we limit our analysis to Solar System orbits
(condition number~\ref{cond:solsis} of Section~\ref{sec:adm_reg}),
because interstellar objects are very rare. As a consequence, we use a
Bayesian theory with a population limited to the Solar System, and all
the probability computations we describe are actually conditional
probabilities to the AR, which is even a subset of the Solar System.

\subsection{Impact probability computation}
Each point on the MOV can be thought as an orbit compatible with the
observations, and we call it Virtual Asteroid (VA). We propagate the
VAs into the future, currently for 30 days from the date of the
observations, and we search for Virtual Impactors (VIs), which are
connected sets of initial conditions leading to an impact
\citep{milani:clomon2}. If a VI has been found on the Modified Target
Plane (MTP, \cite{milani99}), it is associated to a subset $\mathcal
V\subseteq S$ of the sampling space, and hence its probability is
given by
\tiny
\begin{equation*}
\mathbb P(\mathcal V) = \int_{\mathcal V}
p_{\mathbf{S}}(\mathbf{s})\,d\mathbf{s}=
\end{equation*}
\begin{equation}\label{eq:IPcomp}
  =\frac{\displaystyle
    \int_{\mathcal V}\exp\left(-\frac{\chi^2(\mathbf{s})}{2}\right) \det
    M_\mu(\bm\rho(\mathbf{s}))\det
    M_\sigma(\bm\rho(\mathbf{s}))\,d\mathbf{s}}
  {\displaystyle\int_{f_{\sigma}^{-1}(K')}
    \exp\left(-\frac{\chi^2(\mathbf{s})}{2}\right)\det
    M_\mu(\bm\rho(\mathbf{s}))\det
    M_\sigma(\bm\rho(\mathbf{s}))\,d\mathbf{s}}.
\end{equation}
\normalsize
If for a given object we find impacting solutions, we assign to the
object an impact flag, which is an integer number related to the
computation of the impact probability. It depends on the impact
probability and on the arc curvature, as shown in
Table~\ref{tab:impact_flag}. An arc has significant curvature if
$\chi^2>10$, where $\chi$ is the chi-value of the geodesic curvature
and the acceleration (as defined in \cite{milani2007}). The impact
flag can take the integer values from $0$ to $4$: $0$ indicates a
negligible chance of collision with the Earth, whereas the maximum
value $4$ expresses an elevated impact risk ($\ge 1 \%$). It is
conceived as a simple and direct communication tool to assess the
importance of collision predictions, and to give the priority for the
follow-up activities.

\begin{table}[h]
  \begin{center}
    \caption{The table shows the conditions on the Impact Probability
      (IP) and on the arc quality to assign the impact flag to an
      object.} \small{
    \bigskip
    \begin{tabular}{c|l} \hline
           {\textbf{Impact flag}} & \textbf{Condition} \\
           \hline
           0 & $IP\le 10^{-6}$\\
           1 & $10^{-6} < IP \le 10^{-3}$\\
           2 & $10^{-3} < IP \le 10^{-2}$\\
           3 & $IP > 10^{-2}$ and no significant curvature\\
           4 & $IP > 10^{-2}$ and significant curvature\\
           \hline
    \end{tabular}
  }
    \label{tab:impact_flag}
  \end{center}
\end{table}

The goal for a system dedicated to imminent impactors is to detect all
the possible VIs down to a probability level of about $10^{-3}$,
called completeness level~\citep{delvigna2018}.
To reach the completeness level of $10^{-3}$ for our system, we can
neglect the terms which are of smaller order of magnitude in
equation~\eqref{eq:IPcomp}. This allows us to consider the VAs with a
$\chi$-value less than 5. If $\chi=5$ then $\exp(-\chi^2/2)\simeq
10^{-5.4}$, and the corresponding term is negligible. Moreover, the
choice $\chi<5$ is valid for the score computation, because we are
interested in a score accuracy of about $10^{-2}$.

%--------------------------%
%         RESULTS          %
%--------------------------%
\section{Results}
\label{sec:results}

We are setting up a service dedicated to the scan of the Minor Planet
Center NEO Confirmation
Page\footnote{\url{http://www.minorplanetcenter.net/iau/NEO/toconfirm_tabular.html}}
(NEOCP). The goal is to identify asteroids as NEOs, MBOs or distant
objects to be confirmed or removed from the NEOCP, and to give early
warning of imminent impactors, to trigger immediately follow-up
observations. The software used to produce these results is a new
version of the OrbFit Software version
5.0\footnote{\url{http://adams.dm.unipi.it/orbfit/}}.

The service involves the following steps, based on the algorithm
presented in Sections~\ref{sec:cobweb} and \ref{sec:PDF}.
\begin{itemize}
\item Scanning of the NEOCP every $2$ minutes. New cases or old cases
  just updated are immediately run.
\item Computation and sampling of the AR using a
  $2$-dimensional representation in the $(\rho, \dot{\rho})$ plane
  with a either grid or a spider web.
\item Computation of the MOV, obtaining a set of VAs.
\item Propagation of the VAs in the future (currently for 30 days).
\item Projection on the Modified Target Plane, searching for VIs.
\item If VIs exist, computation of the Impact Probability.
\item Computation of the score.
\end{itemize}

The time required to run one target strictly depend on the
characteristics of the object, but usually it is between $15$ and $20$
minutes. When predicting possible imminent impacts, one of the most
important requirements to fulfil is to minimize the number of
unjustified alarms. We mark as non-significant cases the objects for
which there are less than 3 observations or the arc length is
less than 30 minutes (see also Sec.~\ref{sec:gaia}), unless there
exists a nominal solution with a geodesic curvature SNR greater than
1. The classification of a case as non-significant does not
mean we skip the computation: we anyway perform all the steps of the
algorithm, and assign the score and the impact flag. Nevertheless,
being non-significant automatically decreases the priority of the
object in case of an alarm.

Unfortunately, these techniques are not enough to remove all the
spurious cases. They usually occur when the astrometry is either known
to be erroneous or noisy, or anyway not reliable. We cannot solve this
problem, and we acknowledge that the astrometric error models based on
large number statistic are not enough to distinguish erroneous and
accurate astrometry in a small sample (see comments in
Section~\ref{sec:conclusions}).

We test our algorithm on the two well known cases of NEAs that have
impacted the Earth few hours after the discovery, namely $2008~TC_{3}$
and $2014~AA$. We have already pointed out that the choice of the
weights is very important in these cases: to be able to compare the
results with \cite{farnocchia2015}, we choose the same
weights. Furthermore, we also select some cases among the objects that
won't impact to also show the importance of the score computation.

\subsection{Graphical representation of the results}
\label{sec:results_graphical}
We use plots showing the AR and its sampling to present
our results. Hereinafter we describe the color code present in our
figures. Concerning the AR, we make use of the following lines.
\begin{itemize}
  \item The red solid line represents the level curve of the
    heliocentric energy equal to $-k^2/(2a_{max})$. Namely, it is the
    outer boundary of the AR, corresponding to the boundary of the region
    defined by condition~\ref{cond:solsis} in Section~\ref{sec:adm_reg}.
  \item The green dashed line shows where the geocentric energy is
    equal to $0$, also taking into account the condition about the
    radius of the Earth sphere of influence, as discussed in
    Section~\ref{sec:adm_reg}
  \item The magenta dashed line (which is parallel to the range-rate
    axis) represents the shooting star limit condition.
  \item The magenta solid lines (which are parallel to the range-rate
    axis) represent different values of the absolute magnitude.
\end{itemize}
We now provide a description of the colors used for the sampling
points. No point is marked if the $4$-dimensional differential
corrections does not converge, because the point does not belong to
the MOV.
\begin{itemize}
  \item The dots are blue if $\chi \le 2$, and green if $2<\chi\le5$.
  \item The dots are black if $\chi > 5$.
  \item In case a VI has been found, we mark with red circles the
    points representing possible impacting orbits.
  \item The orange star represents the point with the minimum $\chi^2$
    value.
\end{itemize}

\subsection{Asteroid $2008~TC_{3}$}
$2008~TC_{3}$ has been discovered by Richard A. Kowalski at the
Catalina Sky Survey on October 7, 2008. The object was spotted 19
hours before the impact, and it is the first body to be observed and
tracked prior to falling on the Earth. After the discovery, hundreds
of astrometric observations were submitted to the Minor Planet
Center (MPC) and these observations allowed the computation of the
orbit and the prediction of the impact. We use the first tracklet
composed by $4$ observations, and then the first two tracklets ($7$
observations) to ascertain whether we could predict the impact.

\begin{figure}[t!]
    \hspace{-0.3cm}
    \subfloat{%
      \includegraphics[width=0.47\textwidth]{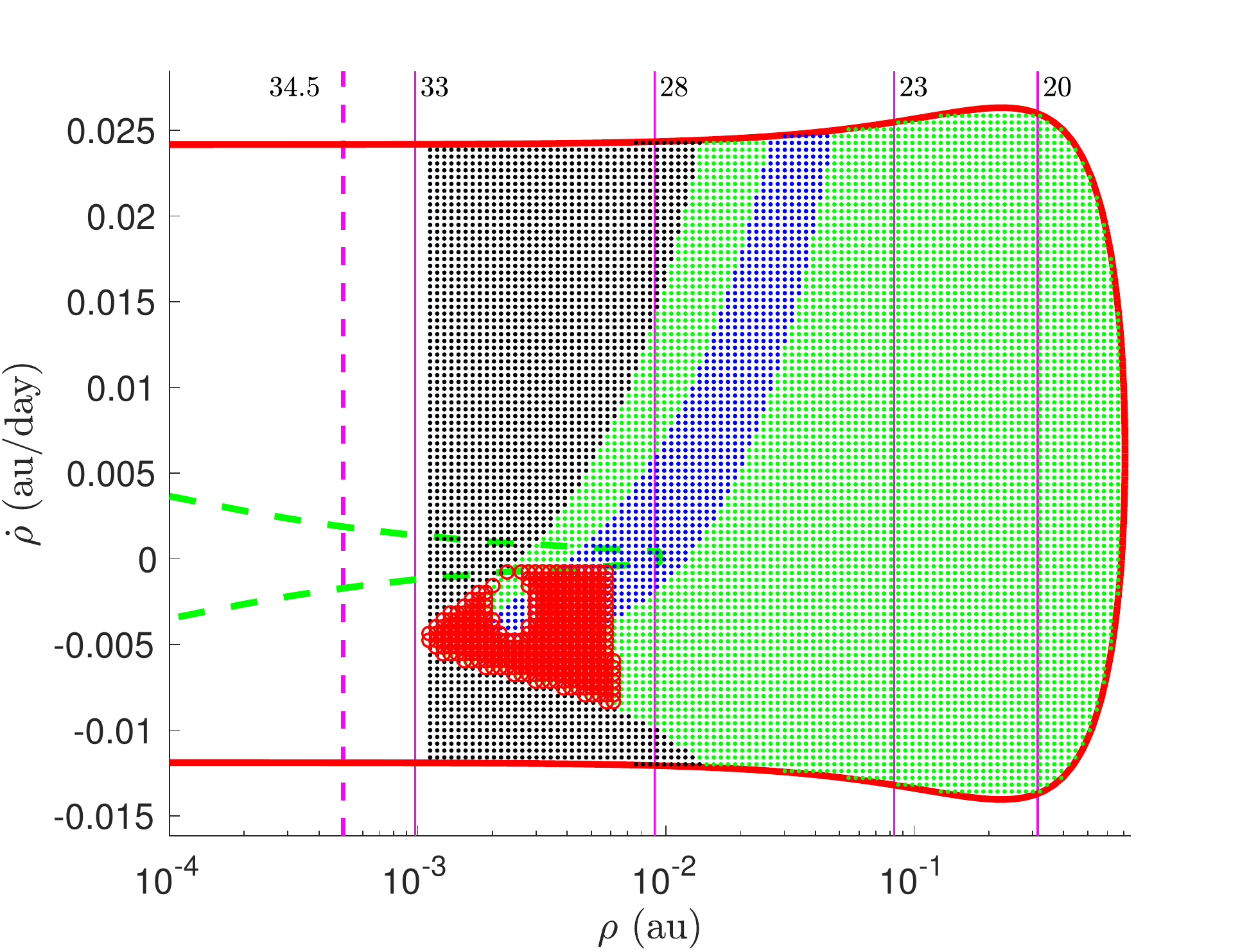}%
    }%
    \hfill%
    \subfloat{%
      \includegraphics[width=0.48\textwidth]{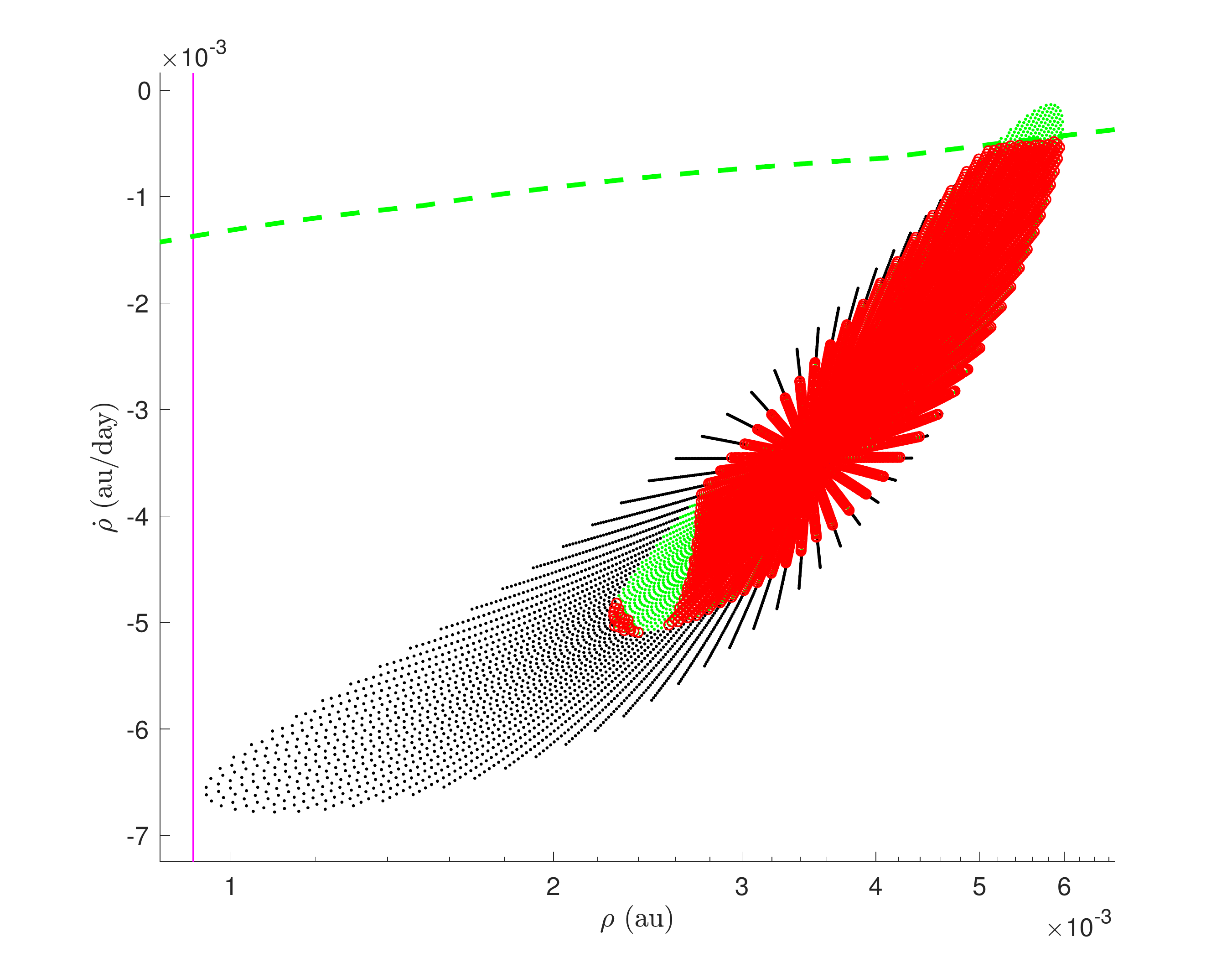}%
    }%
    \caption{Grid sample of the $(\rho,\dot{\rho})$ space for the first
    $4$ observations (top panel) and spider web for the first $7$
    observations (down panel) of $2008~TC_3$.}
    \label{fig:2008TC3_att}
\end{figure}

We compute a uniform densified grid in $\log_{10}(\rho)$
(Fig.~\ref{fig:2008TC3_att}, top panel) where we consider only the
first $4$ observations, while we are able to compute a reliable
nominal orbit, and the consequent spider web using $7$ observations
(Fig.~\ref{fig:2008TC3_att}, down panel). Table~\ref{tab:ip} shows
that with $4$ observations and using the grid we are able to predict a
possible impact of the object with the Earth with an impact
probability of $\simeq 3.6\%$, and the score of the object to be
classified as a NEA is $100\%$. This would have produced an alert for
the observers that could have immediately followed-up the object. With
$7$ observations we can confirm the certainty that the asteroid is a
NEA (score $=100\%$), and the impact probability grows to $99.7\%$.

\subsection{Asteroid $2014~AA$}
$2014~AA$ has been discovered by Richard A. Kowalski at the Catalina
Sky Survey on the new year's eve of $2014$. The object was discovered
21 hours before the impact, but it has not been followed-up as
$2008~TC_3$ because of the exceptional night in which it has been
spotted. We use first the first tracklet composed by $3$ observations,
and then the whole set of $7$ observations to test whether we could
have predicted the impact with our method.

\begin{figure}[h!]
    \hspace{-0.3cm}
    \subfloat{%
      \includegraphics[width=0.47\textwidth]{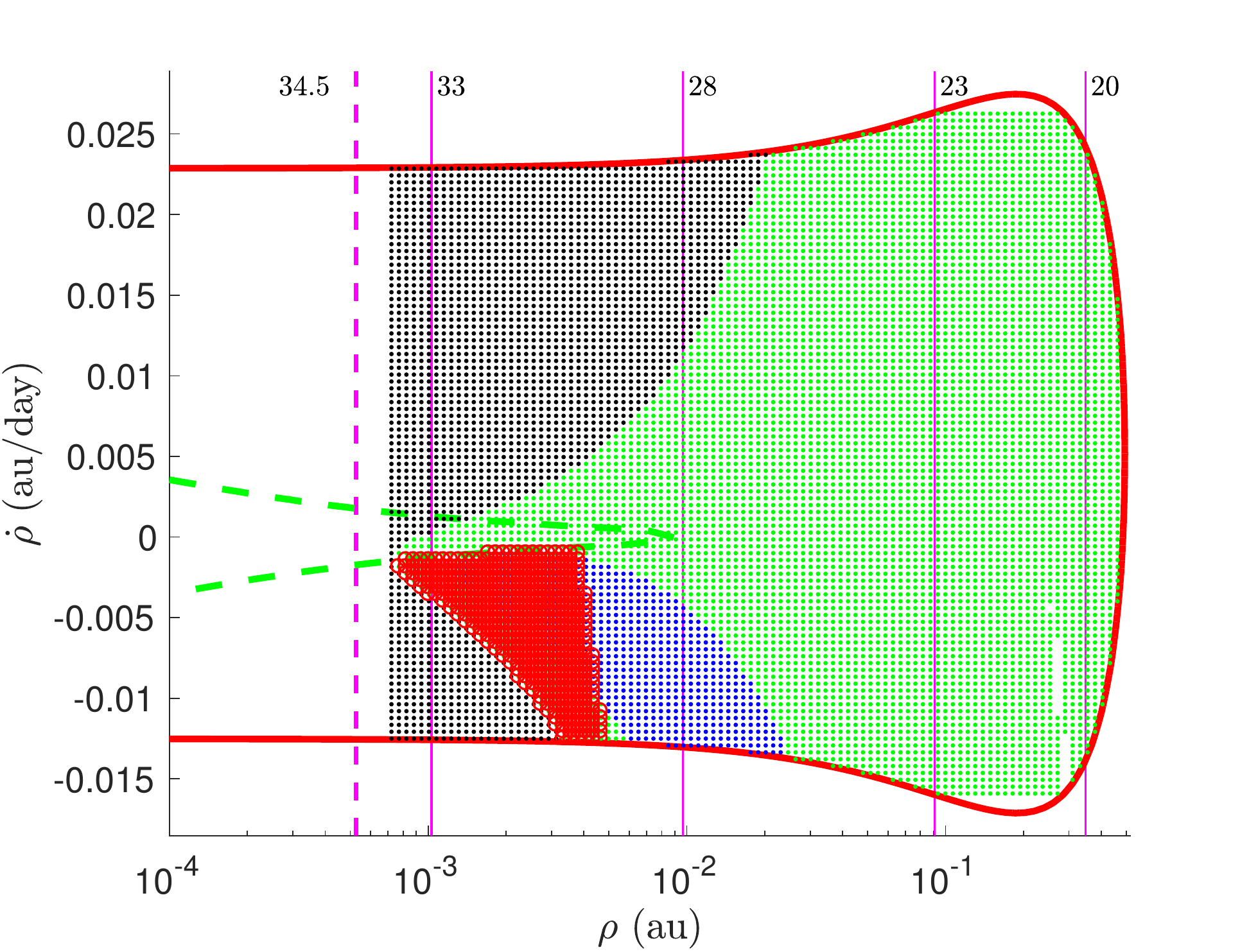}%
    }%
    \hfill%
    \subfloat{%
      \includegraphics[width=0.48\textwidth]{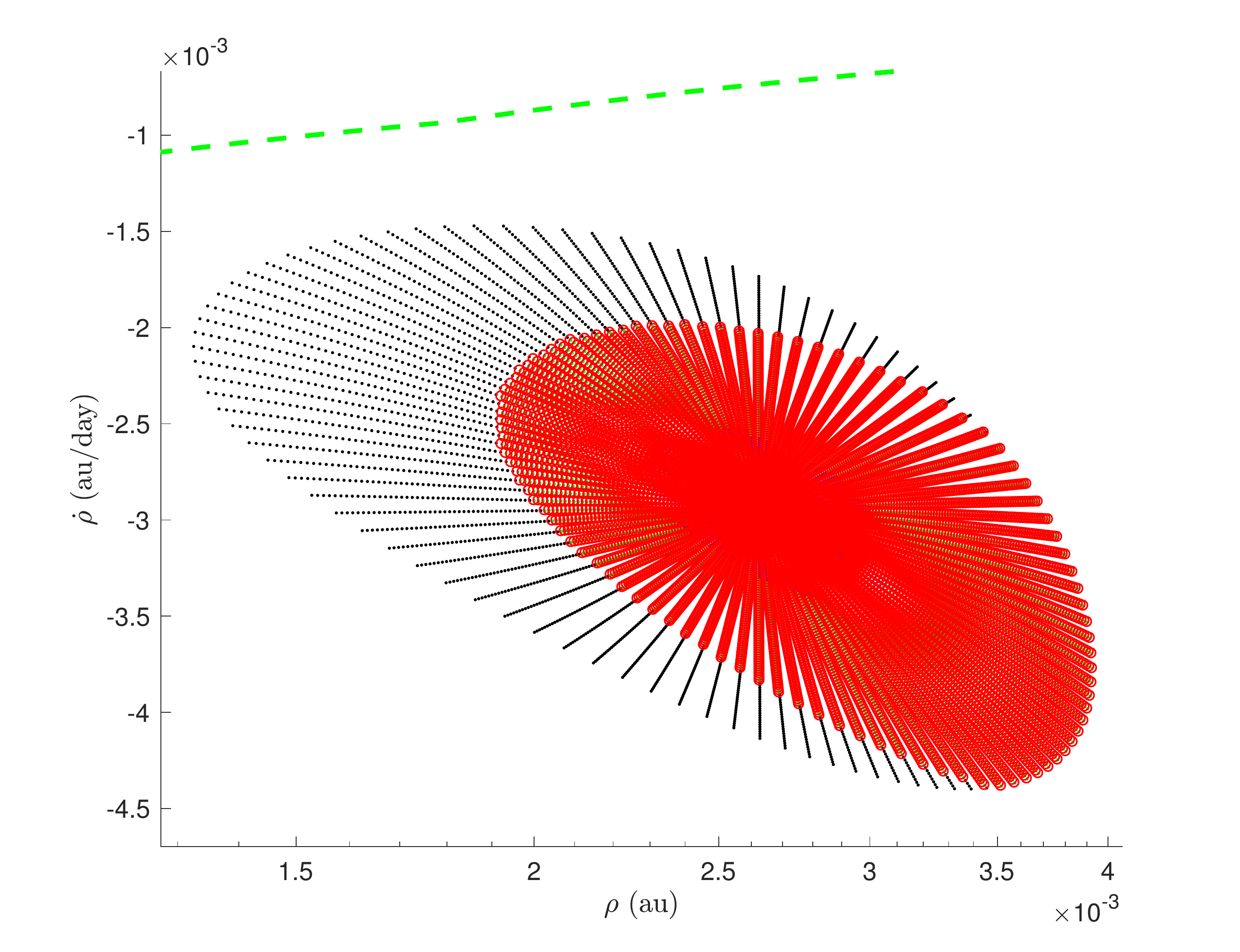}%
    }%
    \caption{Grid sample of the $(\rho,\dot{\rho})$ space for the first
      $3$ observations (top panel) and spider web for the whole set of
      $7$ observations (down panel) of $2014~AA$.}
    \label{fig:2014AA_att}
\end{figure}

\begin{table}[h!]
  \begin{center}
    \caption{The table shows the results of the systematic ranging
      applied to $2008~TC_3$ and $2014~AA$. The columns contain the
      name of the object, the number of observations used, the time
      span covered by the observations, the characteristic of the
      sampling used to compute the MOV (grid or spider web), the score
      of the object (NEA, MBA or Distant), and the impact probability
      (IP).}
    \resizebox{\columnwidth}{!}{\begin{tabular}{l|c|c|c|rrr|r}
        \hline
        \multirow{2}*{\textbf{Name}} & \multirow{2}*{\textbf{$\#$ Obs.}} & \textbf{Time span}& \textbf{Sampling}      &               &\textbf{Score}    &                  & \multirow{2}*{\textbf{IP}}     \\
        &                                                                      & \textbf{(min)}   & \textbf{Grid/Spider}   & \textbf{NEO}    &\textbf{MBO}  & \textbf{DO} &                         \\
        \hline
        $2008~TC_3$   &  4              &    43             &  $\log_{10}(\rho)$ -  grid &  $100\%$        &  0               &      0           &  $3.6\%$\\
        $2008~TC_3$   &  7              &    99             &  Spider                   &  $100\%$        &  0               &      0           &  $99.7\%$\\
        $2014~AA$     &  3              &    28             &  $\log_{10}(\rho)$ -  grid &  $100\%$        &  0               &      0           &  $3.0\%$\\
        $2014~AA$     &  7              &    28             &  Spider                   &  $100\%$        &  0               &      0           &  $100.0\%$\\
        \hline
    \end{tabular}}
    \label{tab:ip}
  \end{center}
\end{table}

These two examples have several analogies: we compute a uniform
densified grid in $\log_{10}(\rho)$ with the first tracklet, that is
only $3$ observations (Fig.~\ref{fig:2014AA_att}, top panel), and we
are able to compute a reliable orbit and the consequent spider web
only with $7$ observations (Fig.~\ref{fig:2014AA_att}, down panel).
Table~\ref{tab:ip} shows that using the first tracklet only, we are
able to predict a possible impact with the Earth with an impact
probability of $\simeq 3.0\%$, and the NEA score of the object is
$100\%$. Again, this would have produced an alert for the observers
that could have immediately followed-up the object. As in the previous
case, with the second tracklet we confirm both that the asteroid is a
NEA (score $=100\%$) and the collision, since the impact probability
grows up reaching the value of $100\%$.

\subsection{Asteroid $2014~QF_{433}$}
The previous examples show how the systematic ranging is capable of
identifying imminent impactors. Although this is one of the most
important applications of this technique, the systematic ranging is
also essential in the first short arc orbit determination
process.

Asteroid $2014~QF_{433}$ has been discovered by F51 - Pan-STARRS 1,
Haleakala on 2014 August 26. The first four observations have been
posted on the NEO Confirmation Page, with the temporary designation
TVPS7NV. It has been on the NEOCP until 2014 September 5. On that day
(with 18 observations) it has been confirmed to be a distant object by the
Minor Planet Center.

\begin{figure}[t!]
    \hspace{-0.3cm}
    \subfloat{%
      \includegraphics[width=0.47\textwidth]{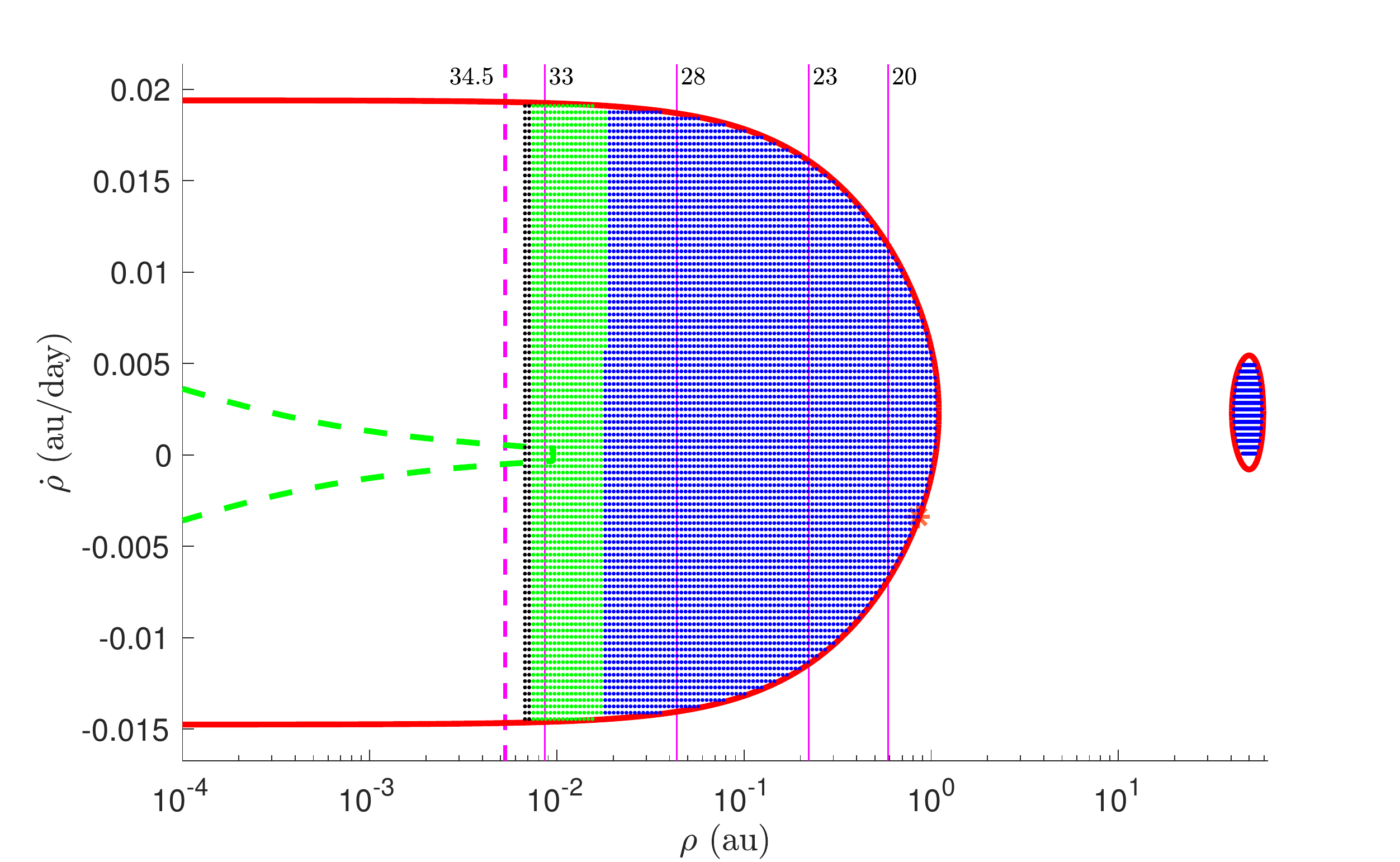}%
    }%
    \hfill%
    \subfloat{%
      \includegraphics[width=0.48\textwidth]{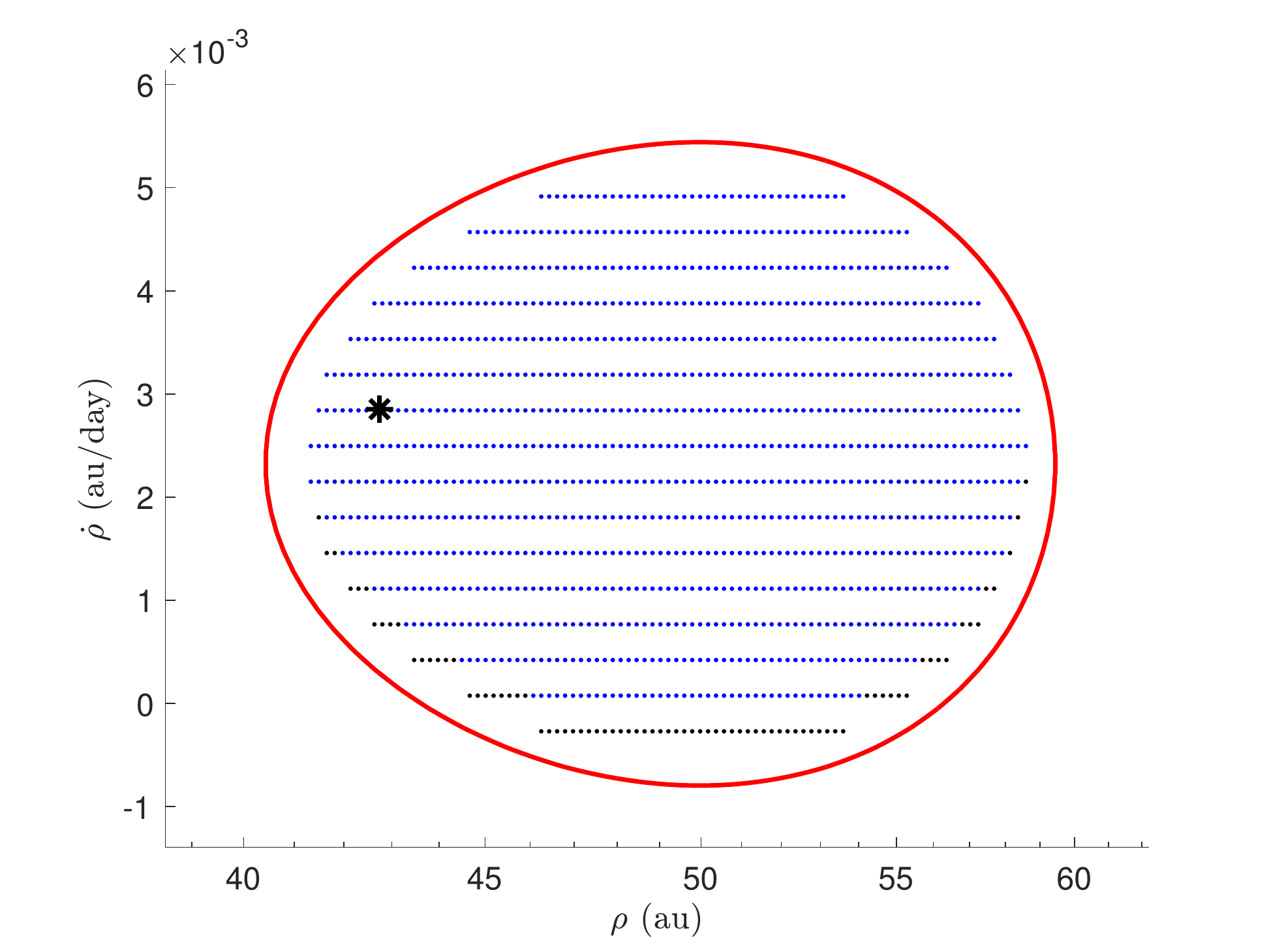}%
    }%
    \caption{Grid sample of the $(\rho,\dot\rho)$ space for the first 4
      observations of $2014~QF_{433}$ (top panel) and an enlargement of
      the second component (down panel). The black star in the
      second component represents the orbit of the object computed with
      all the available observations.}
    \label{fig:TVPS7NV}
\end{figure}

Figure~\ref{fig:TVPS7NV} shows the results of the systematic ranging
on this asteroid, with the four discovery observations only, and 51
minutes of arc length. In this case the AR has two connected
components, indicating the possibility for the object to be distant. The
values of the three positive roots of equation of degree $6$ are
$r_1=1.103$ au, $r_2=40.072$ au, and $r_3=59.786$ au. The
attributable is
\tiny
\[
\mathcal A = (\alpha,\delta,\dot\alpha,\dot\delta)=
(5.7358902, -0.3008327, -3.35275\cdot 10^{-4}, -9.94065\cdot 10^{-5}),
\]
\normalsize
with $\alpha$ and $\delta$ in radians and $\dot\alpha$ and
$\dot\delta$ in radians per day.  The two plots in
Figure~\ref{fig:TVPS7NV} clearly show that the object is distant,
since almost all the grid points corresponding to the MOV lie in the
second connected component. As a consequence, the cumulative score for
the Distant and Scattered classes is $99\%$.

As a further validation, we take the orbital elements of this asteroid
from the AstDyS database\footnote{Asteroid Dynamics Site, available at
  \url{http://hamilton.dm.unipi.it/astdys/}}, and we compute the range
and the range-rate at the epoch of the attributable. The result is
shown in the down panel of Figure~\ref{fig:TVPS7NV}: the black star
represents the orbit of $2014~QF_{433}$ computed with all the
available observations, and it is in perfect agreement with the
systematic ranging sampling.

\subsection{Asteroid $2017~AE_{21}$}
The case of $2017~AE_{21}$ shows the importance of the score
computation. An object is worthy of attention and has to be
followed-up even though it is not an impactor: for instance, it could
be a potential NEA.

\begin{figure}[t!]
    \hspace{-0.3cm}
    \subfloat{%
      \includegraphics[width=0.47\textwidth]{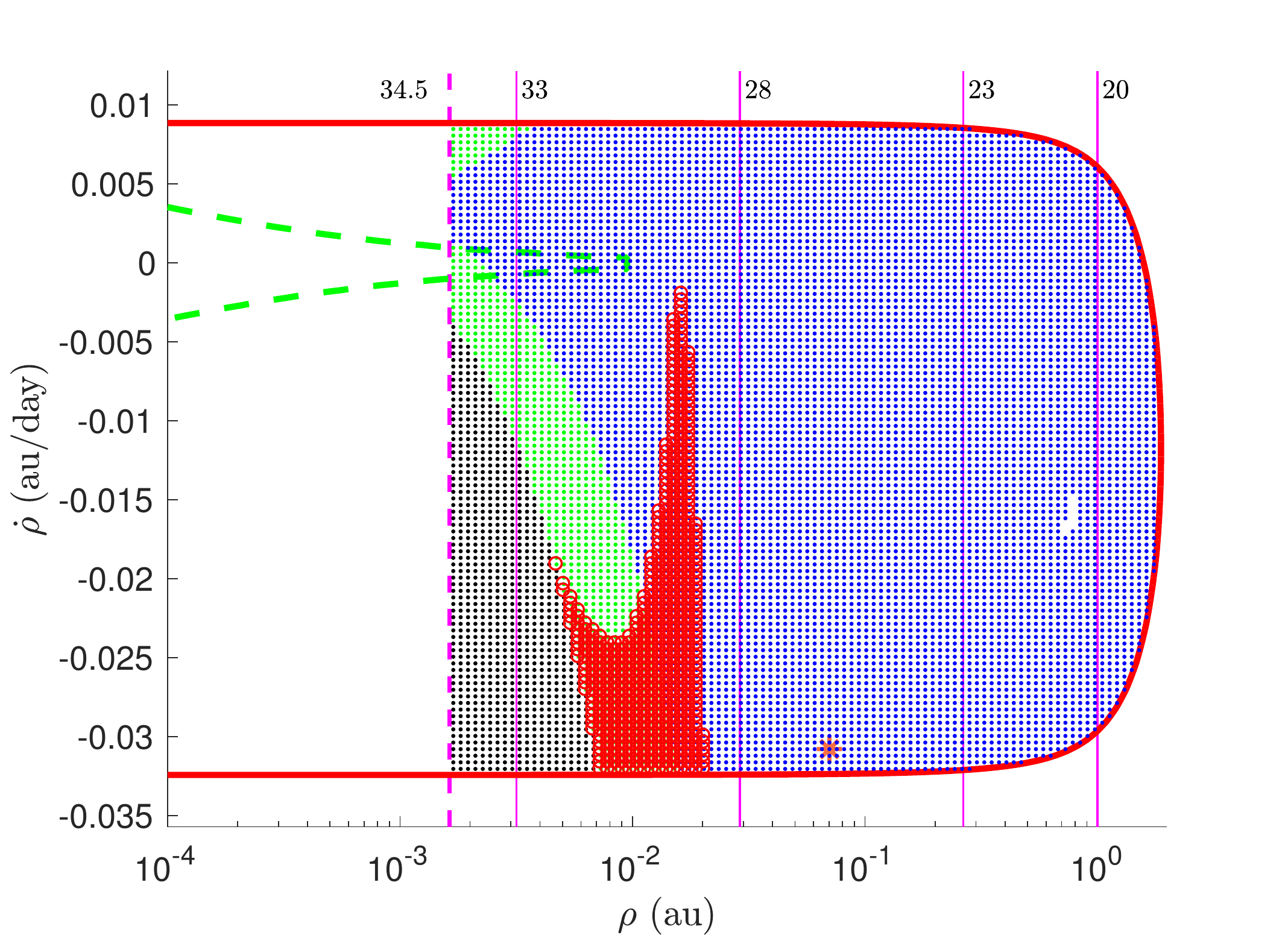}%
    }%
    \hfill%
    \subfloat{%
      \includegraphics[width=0.48\textwidth]{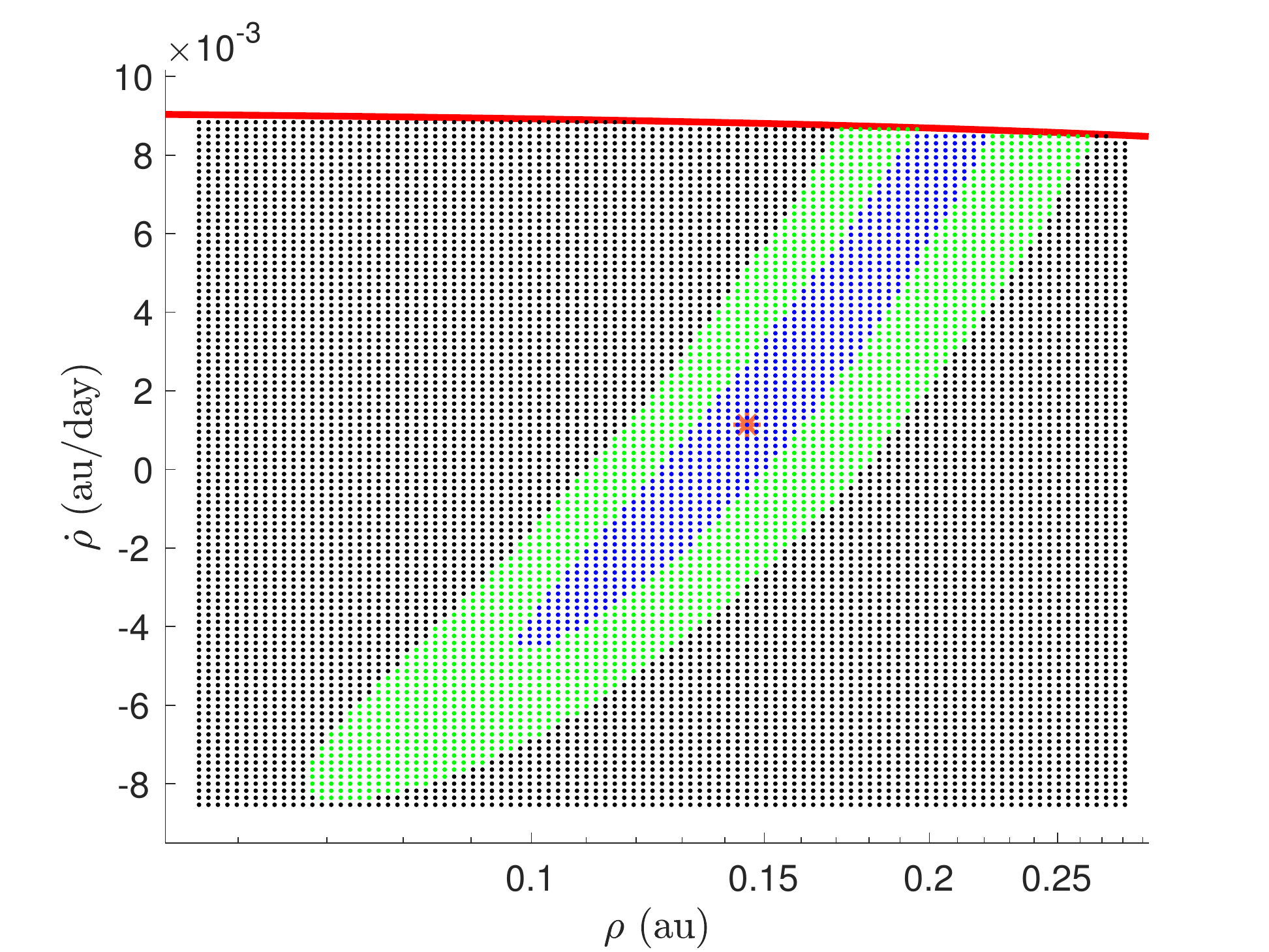}%
    }%
    \caption{Grid sample of the $(\rho,\dot{\rho})$ space for the first
      $3$ observations (top panel) and for the whole set of $5$
      observations (down panel) of $2017~AE_{21}$. In both cases we do
      not have any reliable nominal orbit to use, and as a consequence
      we adopt the grid sampling.}
    \label{fig:P10yBuc}
\end{figure}

Asteroid $2017~AE_{21}$ has been discovered by F51 - Pan-STARRS 1,
Haleakala on 2017, January 3. It appeared on the NEOCP as a tracklet
of 3 observations spanning 30 minutes, with the temporary designation
P10yBuc. It has been confirmed to be a NEA on 2017 January 24, when it
had 5 observations. With the first tracklet, our system produces an
impact flag of 2, indicating a modest impact risk, with an impact
probability $IP = 2\cdot 10^{-3}$. Moreover, the NEO score is $92\%$,
encouraging some follow-up to confirm. The top panel of
Figure~\ref{fig:P10yBuc} shows the result when using the first
tracklet only. We do not have any reliable nominal orbit to use, and
as a consequence we adopt the grid sampling. The portion of the grid
corresponding to low $\chi$ values (blue points) is quite wide,
indicating a great uncertainty in the orbit determination, and the
uncertainty region contains also impacting solutions.

With just two additional observations, the differential corrections
still fail in computing a reliable nominal orbit, but now the good
portion of the grid is located in a small subregion of the AR (see
Figure~\ref{fig:P10yBuc}, down panel). In this case the uncertainty
region does not contain impacting orbits, thus we get an impact flag
of 0, with $IP = 0$, whereas the NEO score grows to $100\%$. As a
consequence, the low probability VI has been contradicted by the new
observation, but the follow-up suggestion coming from the high $98\%$
NEO score of the first run was reliable.

\subsection{NEOCP object $P10vxCt$}
As we stated in the introduction of this section, noisy astrometry can
be the cause for unjustified alarms. In fact, if an object has a
single tracklet of few observations and one of them is erroneous, the
arc usually shows a significant curvature, implying that the object
seems very close and fast moving. Most likely, it could be classified
as an immediate impactor with very high impact probability.

\begin{figure}[t!]
    %\hspace{-0.3cm}
    \subfloat{%
      \includegraphics[width=0.45\textwidth]{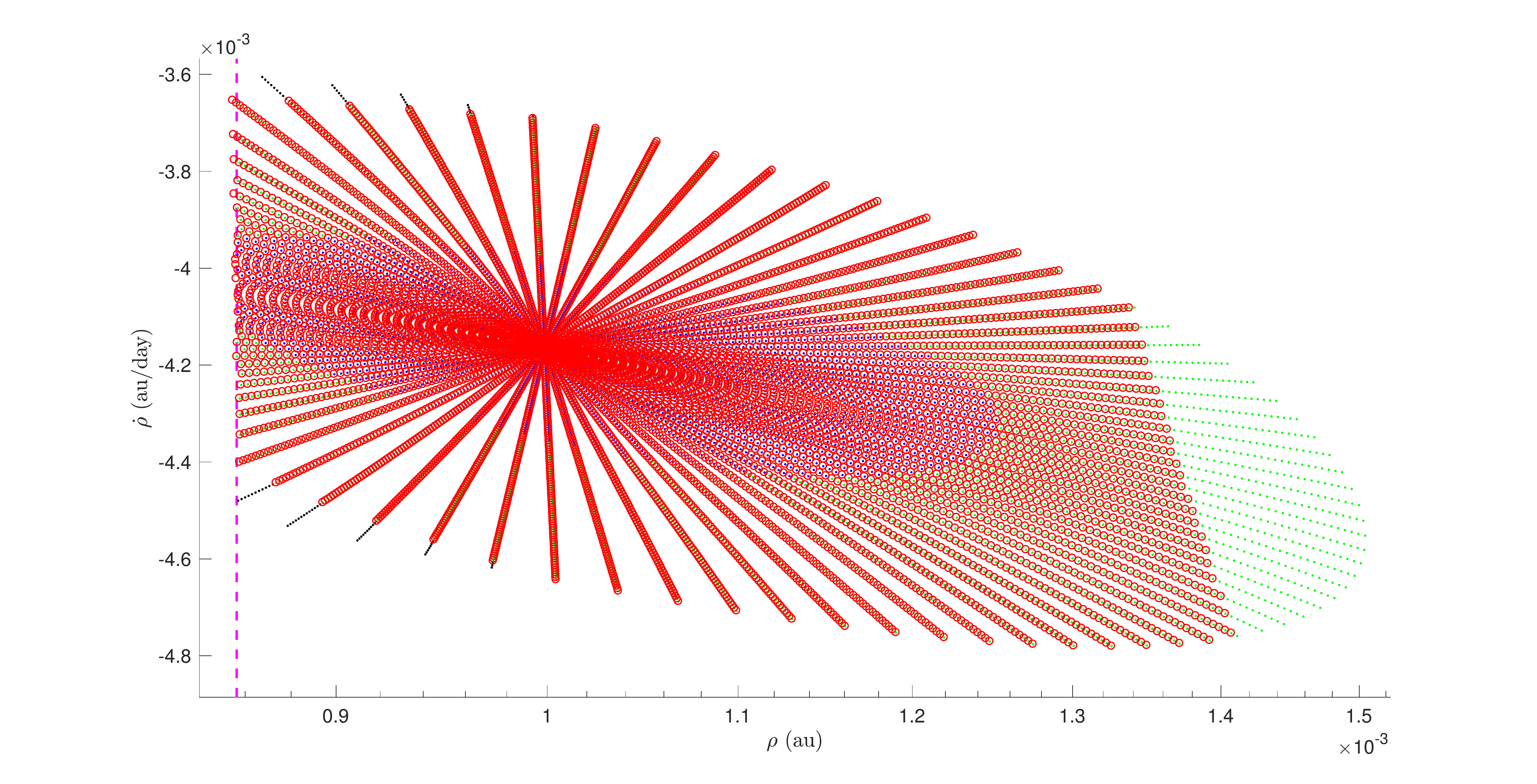}%
    }%
    \hfill%
    \subfloat{%
      \includegraphics[width=0.44\textwidth]{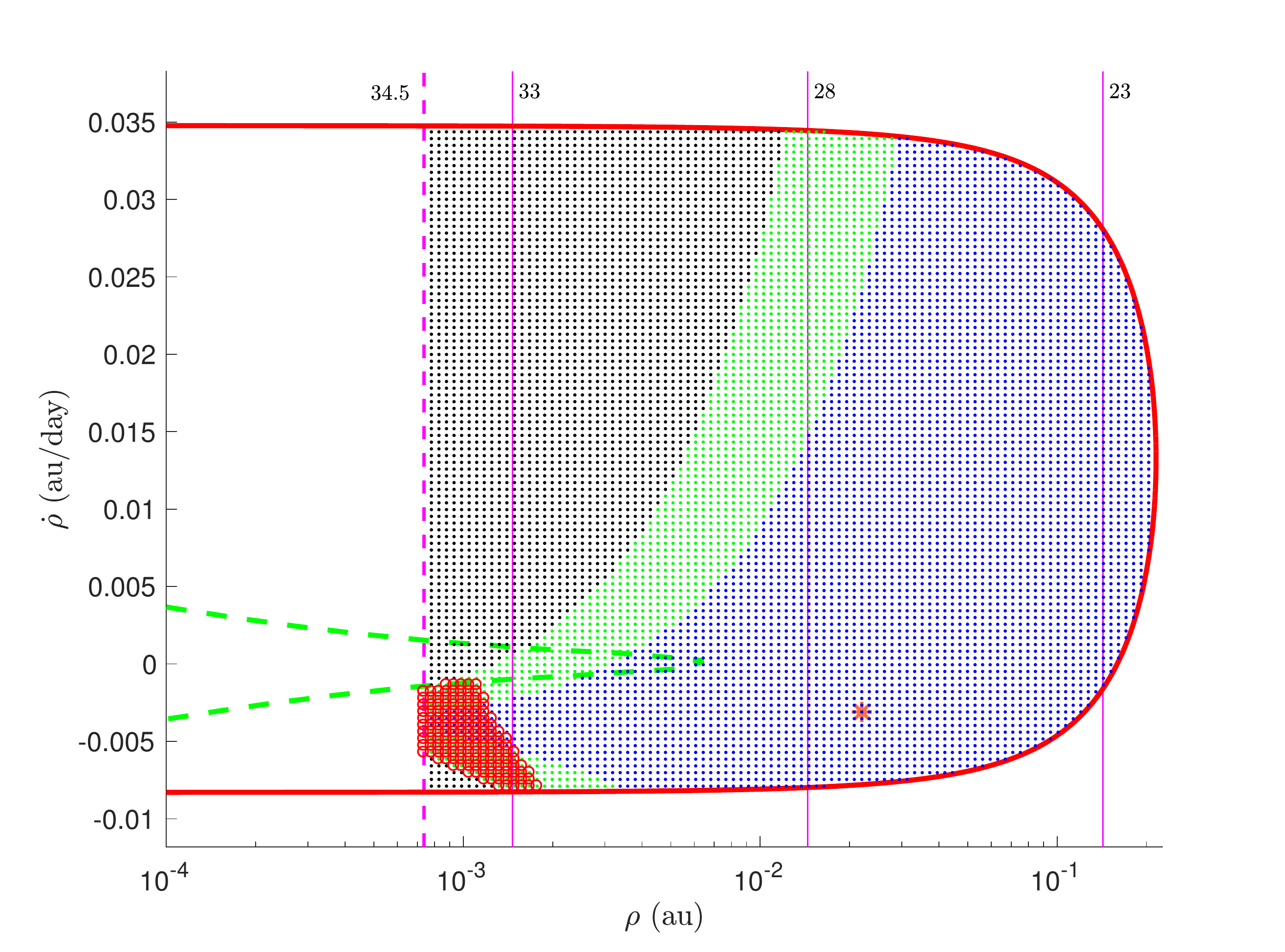}%
    }%
    \caption{Grid sample of the $(\rho,\dot{\rho})$ space for the first
      $3$ observations (top panel) and for their remeasurement (down
      panel) of $P10vxCt$.}
    \label{fig:P10vxCt}
\end{figure}

Object $P10vxCt$ has been spotted by F51 - Pan-STARRS, Haleakala on
2016 June 8. The first time it appeared on the NEOCP it had a tracklet
with 3 observations spanning about 44 minutes
(Table~\ref{tab:P10vxCt}, above). It has never been confirmed, but it
is anyway an important example to show the risk posed by noisy
astrometric data. With the first 3 observations, our system computes a
nominal solution compatible with a very close orbit, resulting in a
spider web sampling over a small subset of the AR (see
Fig.~\ref{fig:P10vxCt}, top panel). A very large fraction of the MOV
orbits are impacting solutions, and it results in an impact
probability of 99.2\% and impact flag 4, considering the significance
of the curvature. The second batch of observations consists of 4
positions: 3 of them are a remeasurement of the first tracklet
obtained from the discovery images of the object, plus an additional
observation. With this new astrometry, the impact has been ruled out
and the object has been removed from the NEOCP.

\begin{table}[t!]
  \begin{center}
    \caption{Astrometric data for NEOCP object $P10vxCt$.  First
      tracklet with 3 observations (above), and remeasurement of the
      first tracklet from the discovery images (below).}
    \vspace{0.5cm}
     \resizebox{\columnwidth}{!}{\begin{tabular}{l c c c l}
      \hline
      \textbf{Date (UTC)} & $\alpha$ & $\delta$ & $R$ & \textbf{Code} \\
      \hline
      2016-06-08.29327 & 13 13 16.962 & $-20$ 25 56.90 & 21.0 & F51\\
      2016-06-08.30357 & 13 13 12.688 & $-20$ 28 31.36 & 20.9 & F51\\
      2016-06-08.32416 & 13 13 04.699 & $-20$ 33 46.35 & 21.0 & F51\\
      \hline
    \end{tabular}
    \label{tab:P10vxCt}
     }\\
    \vspace{0.5cm}
    \resizebox{\columnwidth}{!}{\begin{tabular}{l c c c l}
      \hline
      \textbf{Date (UTC)} & $\alpha$ & $\delta$ & $R$ & \textbf{Code} \\
      \hline
      2016-06-08.293273 & 13 13 16.963 & $-20$ 25 56.53 & 20.3 & F51\\
      2016-06-08.303571 & 13 13 12.856 & $-20$ 28 33.24 & 20.5 & F51\\
      2016-06-08.324159 & 13 13 04.683 & $-20$ 33 46.33 & 20.4 & F51\\
      \hline
    \end{tabular}}
  \end{center}
\end{table}

\begin{figure}[h]
  \centering
  \includegraphics[width=0.9\columnwidth]{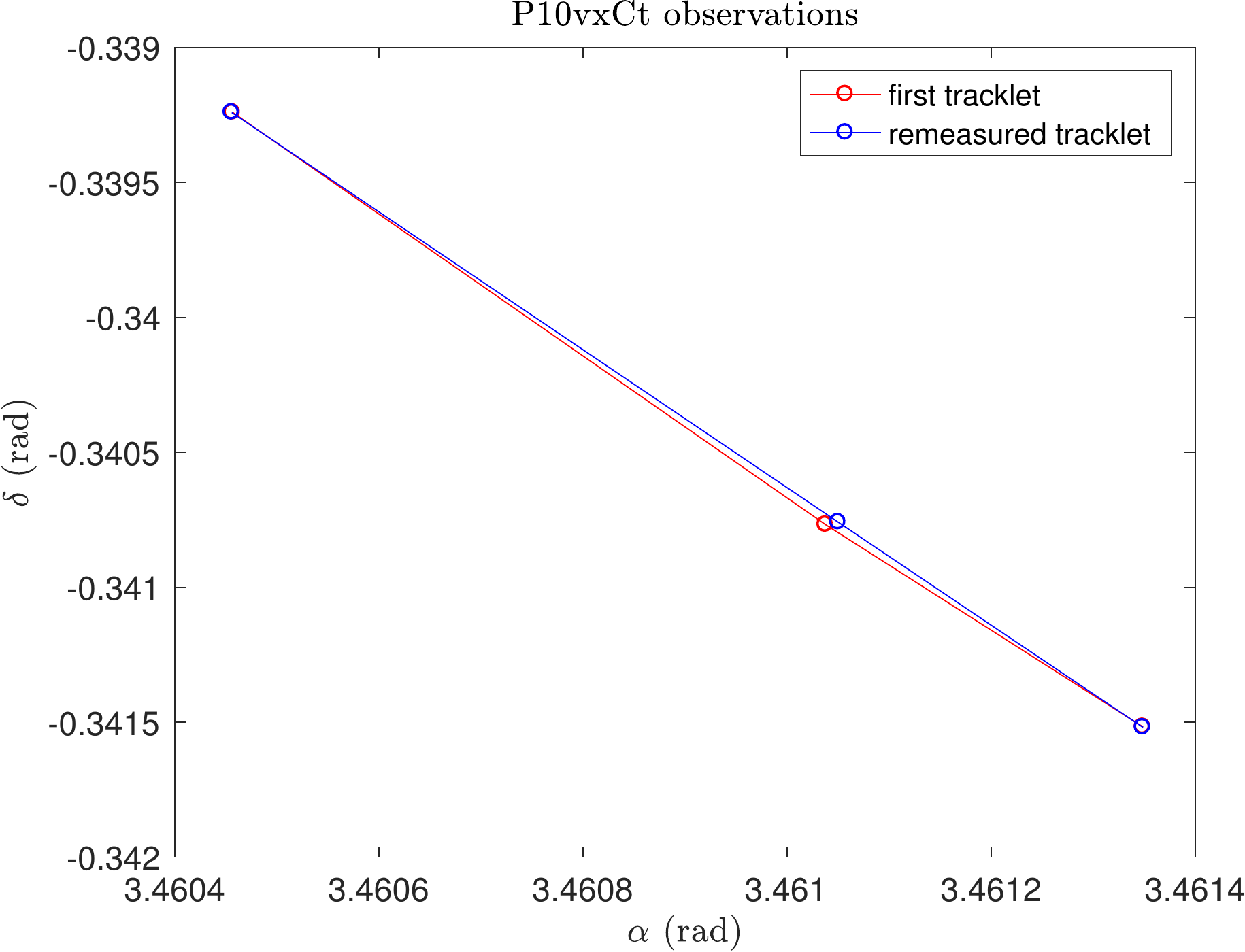}
  \caption{Plot of the two batches of observations reported in
    Table~\ref{tab:P10vxCt}, for NEOCP object $P10vxCt$. The red line
    represents the originally submitted tracklet, whereas the blue one
    the remeasured tracklet. The higher curvature of the first arc
    with respect to the second is clear.}
  \label{fig:curv_P10vxCt}
\end{figure}

To show the role of the remeasurements in the impact removal, we
consider the 3 remeasured observations only (see
Table~\ref{tab:P10vxCt}, below). The second observation in the first
tracklet was badly determined, being off by about $3$ arcsec from the
corresponding one in the second batch. The effect of this shift can be
seen in the curvature parameters $\kappa$ (geodesic curvature) and
$\dot\eta$ (acceleration). For the first tracklet we have
\tiny
\[
\kappa_1 = (0.0010073\pm 0.0001015)\quad\text{and}\quad \dot\eta_1
= (0.0003218\pm 0.0001013),
\]
\normalsize
whereas for the remeasured tracklet
\tiny
\[
\kappa_2 = (0.0000649\pm 0.0006749)\quad\text{and}\quad \dot\eta_2
= (-0.0000430\pm 0.0006750),
\]
\normalsize
both significantly lower than the first ones (see
Fig.~\ref{fig:curv_P10vxCt} for a graphic representation of the two
arcs). Moreover, as we can see from the curvature uncertainties, both
curvature components are not significantly different from $0$. As a
consequence, with the remeasured observations only, the impact
solution is sharply downgraded: a nominal solution cannot be computed
anymore, resulting in a grid sampling of the AR, and the impact orbits
are a very small fraction of the MOV orbits (see
Fig.~\ref{fig:P10vxCt}, right panel). Thus the impact probability
lowers to about $IP = 7.5\cdot 10^{-5}$, with an impact flag of 1.

Providing remeasured observations is not the only way to solve the
problem caused by bad astrometry. The second observation is not as
good as the other two, and let us suppose this information were
provided along with the observation itself. In this case, we could
have properly down-weighted the second observation to take into
account the additional information, and the case would have been
solved. To prove this claim, we assign a formal uncertainty of $3$
arcsec to both the right ascension and the declination of the second
observation. With this choice, the impact solution still remain, but
with an impact probability $IP=4.4\cdot 10^{-4}$. Until this
additional metadata will be provided together with the observations,
cases like the one presented here can be solved only by a manual
intervention after all the computations (remeasurement) or by a fast
follow-up (see Section~\ref{sec:conclusions} for general comments on
this issue).

%---------------------%
%        GAIA         %
%---------------------%
\section{The ESA Gaia mission and short arc orbit determination}
\label{sec:gaia}
The ESA Gaia mission, currently surveying the sky from the Sun-Earth
$L_2$ Lagrangian Point, is providing astrometry of stars and
asteroids, at the sub-milliarcsec accuracy~\citep{prusti2012} down to
magnitude $V=20.7$. The spin of the satellite is $6$ h, and it
operates in a continuous scanning mode.  It has two lines of sight,
separated by an angle of $106.5^{\circ}$ in its scanning
direction. The continuous mode of observation implies targets are not
pointed at, but are rather passing in front of Gaia fields of
view. Such crossings are called transits. Over $5$ years of nominal
mission duration, the objects observed by Gaia will have a coverage of
$80-100$ observations for an average direction ($60-70$ for the
ecliptic~\citep{tanga2016})

Gaia focal plane is a large Giga-pixel array of 106 CCDs. The CCDs are
organized as follows.
\begin{itemize}
\item The first two CCD strips are devoted to the source
  detection. This is the instrument called Sky Mapper (SM).
\item The following 9 strips are astrometric CCDs (Astrometric Field,
  AF).
\item Other CCD strips are devoted to low resolution
  spectro-photometry (red and blue photometry, RP/BP) and high
  resolution spectroscopy (RVS), which is not considered for asteroid
  studies.
\end{itemize}

Each Solar System Object (SSO) transit is composed, at most, by 10
astrometric observations (AF and SM instruments), distributed
over 50 seconds.  The Data Processing and Analysis Consortium (DPAC)
has, as part of its activities, the source identification and further
processing done on the ground. In this context, the Coordination Unit
4 (CU4) performs the analysis of objects deserving a specific
treatment, as Solar System object (PI: P. Tanga). The DPAC CU4 has
implemented two pipelines for Solar System processing
~\citep{tanga2007, mignard2007}:

\begin{itemize}
\item SSO-ST is the Solar System short-term processing, devoted to
  provide a first, approximate orbit for the recovery of
  objects potentially discovered by Gaia. A ground-based
  follow-up network (Gaia-FUN-SSO,~\cite{thuillot2016}) is currently
  operating, realizing follow-up observations of Gaia
  potential discoveries from the ground;
\item SSO-LT is the Solar System long-term processing, which runs
  for the data releases, performing a more sophisticated data
  reduction with the best possible instrument calibration and
  astrometric solution.
\end{itemize}

\subsection{The short-term processing}
\label{sec:Gaia_ST}
The Solar System short-term processing is based on few transits of the
object (at least $3$), covering a time span of few hours (at
  least six). At now, the astrometry for the alerts is based on a
preliminary calibration, which is not the same used for the long-term
processing. As a result, the error model required by these
observations is not different from the one already used for the
ground-based cases with the best astrometry. We uniformly weight
($0.1$ arcsec) both right ascension and declination.

If the detected source is not successfully linked with a known solar
system object, then it is potentially a discovery. It is thus crucial
to predict a possible sample of orbits for the ground-based follow-up
network to certify the discovery. In the framework of the CU4 data
treatment, this is done by random-walk statistical ranging, which has
been developed by \cite{muinonen2016}. This is the Java code which is
currently producing the orbital data for Gaia alerts in the SSO-ST
pipeline.

We take the opportunity of the availability of the method presented in
Sec.~\ref{sec:sampling} to validate independently the results of the
SSO-ST pipeline.

The impact probability computation, which is a key feature when we
look for possible impactors, is not so essential when we have to deal
with Gaia short-term observations. We expect indeed that among the
objects that Gaia will discover, there will be a large fraction of
Main Belt Asteroids, few Near Earth Asteroids, and it would be very
surprising if it could discover even an imminent
impactors~\citep{carry2014_gfs}. On the other side, the use of the
double grid or of the cobweb is essential in this case.

\subsection{Results}
The same service presented in Section~\ref{sec:results} to scan the
NEO Confirmation Page, has been ran on possible Gaia
discoveries. The graphical representation of the results is identical
to the one given in Section~\ref{sec:results_graphical}. We also
use the same version of the OrbFit software cited in
Sec.~\ref{sec:results}, adapted to deal with a different input given
by Gaia observations.

All the alerts are available at \url{https://gaiafunsso.imcce.fr/},
and hereinafter we present two particular objects, that at the time
seemed to be possible Gaia discoveries, but then their observations
have been linked to ground-based observations submitted earlier in
time.

The first object that could have been discovered by Gaia on 2016,
December 29, is a Main Belt Asteroid. It has been identified as
g0T015. It has four Gaia transits which cover a time span of
$\simeq 16$ hours.

\begin{figure}[t!]
  %\hspace{-10mm}
    \subfloat{%
      \includegraphics[width=0.44\textwidth]{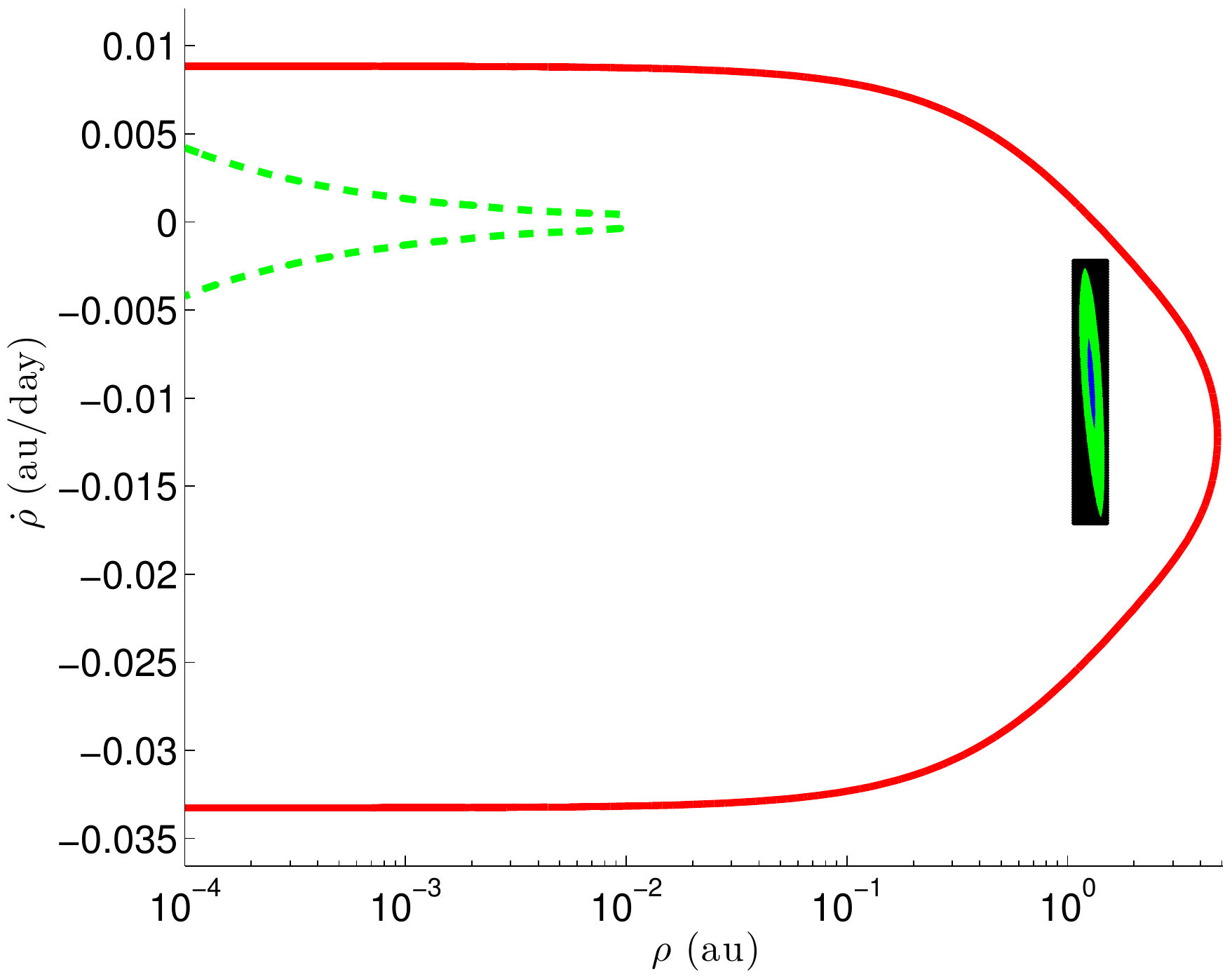}%
    }%
    \hfill%
    \subfloat{%
      \includegraphics[width=0.45\textwidth]{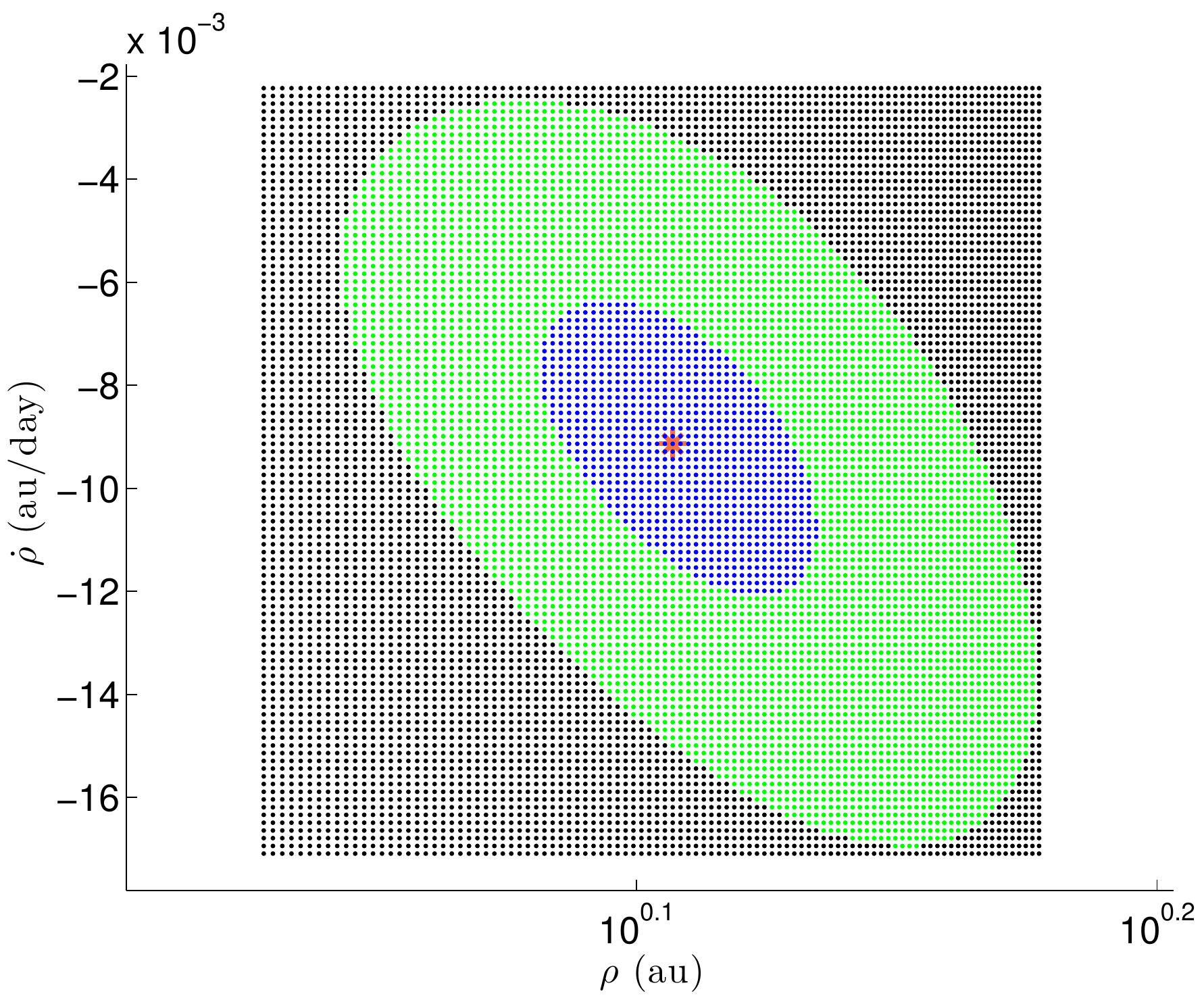}%
    }%
    \caption{Top panel: grid sample of the $(\rho,\dot{\rho})$ space
      for the Gaia object $g0T015$. Down panel: zoom of the second
      grid.}
    \label{fig:g0T015_AR}
\end{figure}

We apply the systematic ranging (using the two grids) on the
observations, and the results are summarized in
Fig.~\ref{fig:g0T015_AR}. The object has been then followed-up by the
Observatoire de Haute-Provence (OHP) for two consecutive nights (2017,
January 3-4), and the observations have been reported to the Minor
Planet Center.  The object has been recognized as a Main Belt Object,
and it obtained the provisional designation $2017~$AD$_{17}$. It could
have been the first Gaia discovery, if it hasn't been linked to few
observations from F51 Pan-STARSS submitted earlier (March
  2014).

To be sure that the observed object was the same potentially
  discovered by Gaia, we consider as initial guess the elements
corresponding to the point with the minimum value of $\chi^2$, and we
perform an orbit determination process including the outlier
rejection procedure~\citep{milani:orbdet}.

\begin{center}
  \begin{figure}[h!]
    \includegraphics[width=\columnwidth]{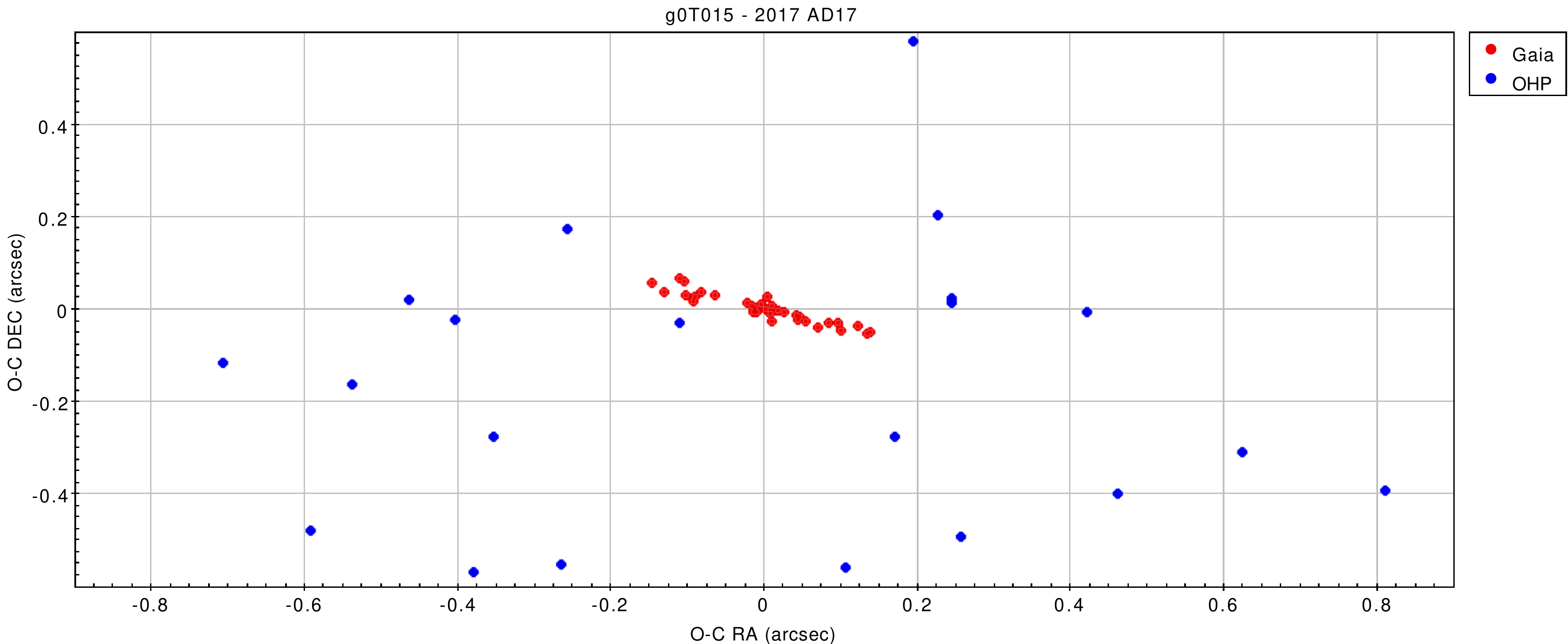}
    \caption{Residuals in right ascension and declination, in arcsec,
      obtained as a result of the orbit determination applied to the
      whole set of observations: Gaia (red points) and ground based
      (blue points) for the Gaia object $g0T015$ ($2017~$AD$_{17}$).}
    \label{fig:g0T015_res}
  \end{figure}
\end{center}

Figure~\ref{fig:g0T015_res} shows that follow-up observations from the
OHP, and Gaia observations match together, according to the orbit
selected as first guess. The residuals for the OHP (blue points) are
way larger than the ones obtained from Gaia observations (red points),
as expected. The weights used in this case are uniform in right
ascension (RA) and declination (DEC), and correspond to $0.1$
arcsec. The result also shows a high correlation (close to 1) between
RA and DEC, which is typical for Gaia observations due to its scanning
law.

\begin{figure}[t!]
    \hspace{-0.3cm}
    \subfloat{%
      \includegraphics[width=0.44\textwidth]{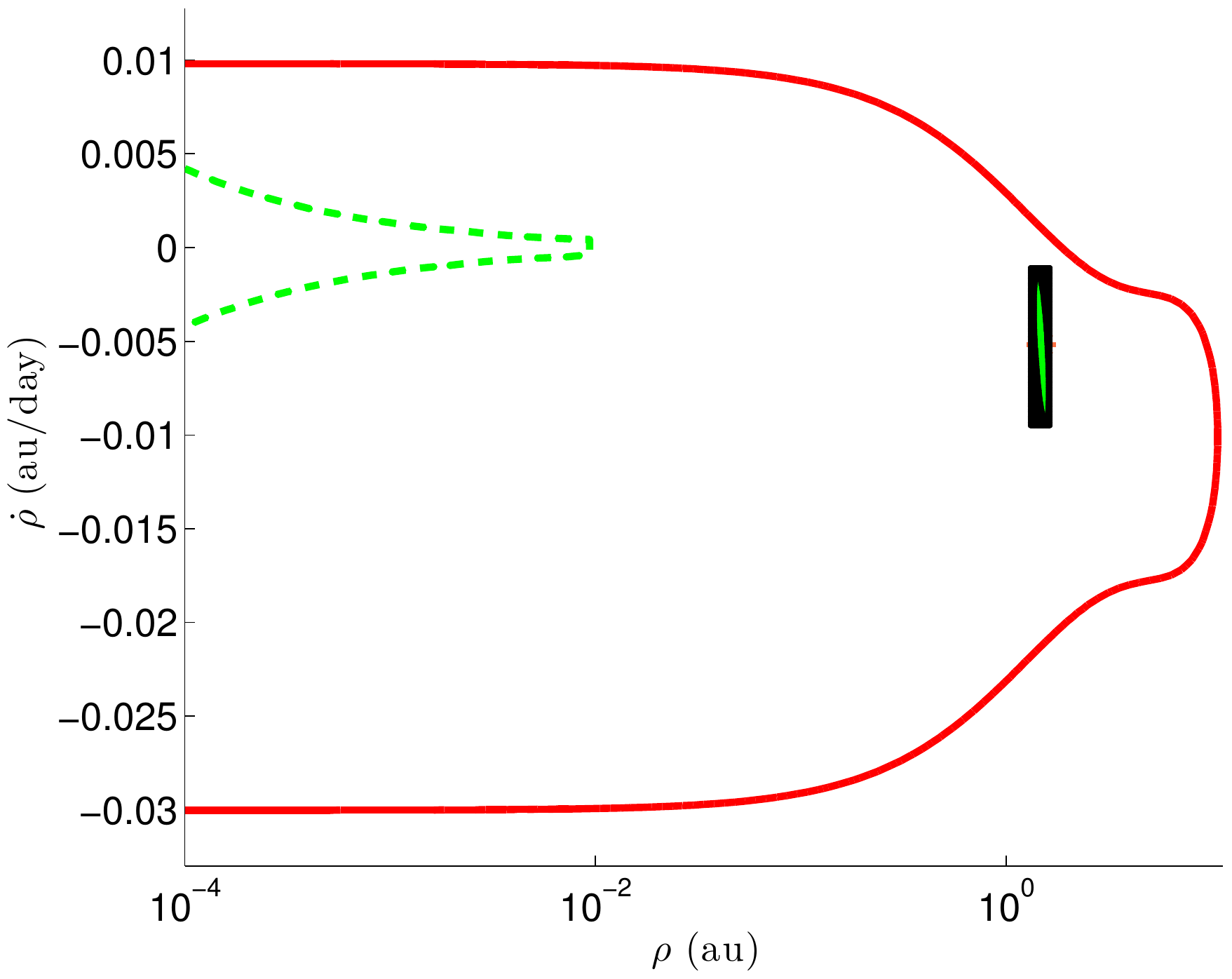}%
    }%
    \hfill%
    \subfloat{%
      \includegraphics[width=0.45\textwidth]{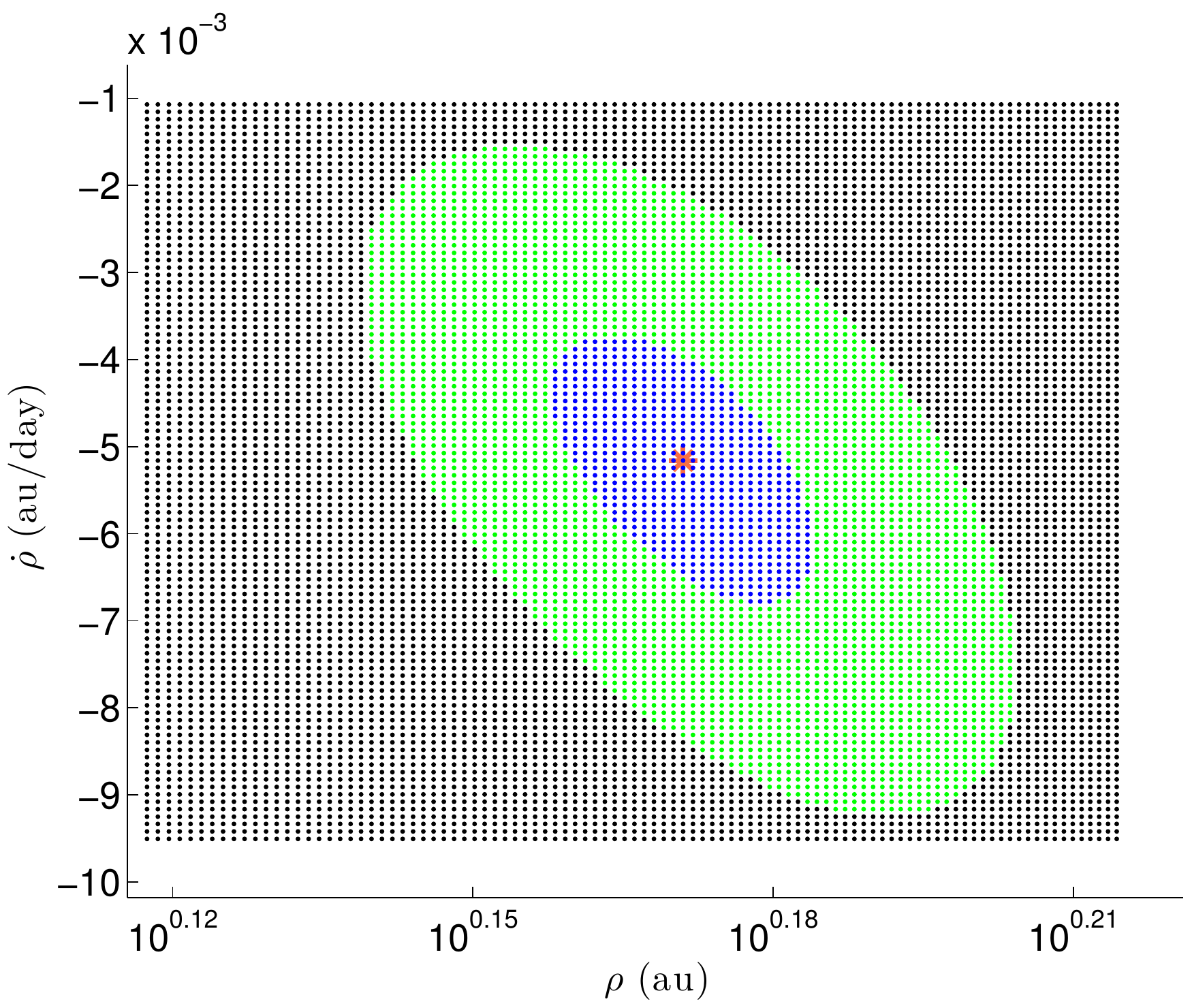}%
    }%
    \caption{Top panel: grid sample of the $(\rho,\dot{\rho})$ space
      for the Gaia object $g1j0D7$. Down panel: zoom of the second
      grid.}
    \label{fig:g1j0D7_AR}
\end{figure}

\begin{center}
  \begin{figure}[h!]
    \includegraphics[width=0.7\columnwidth]{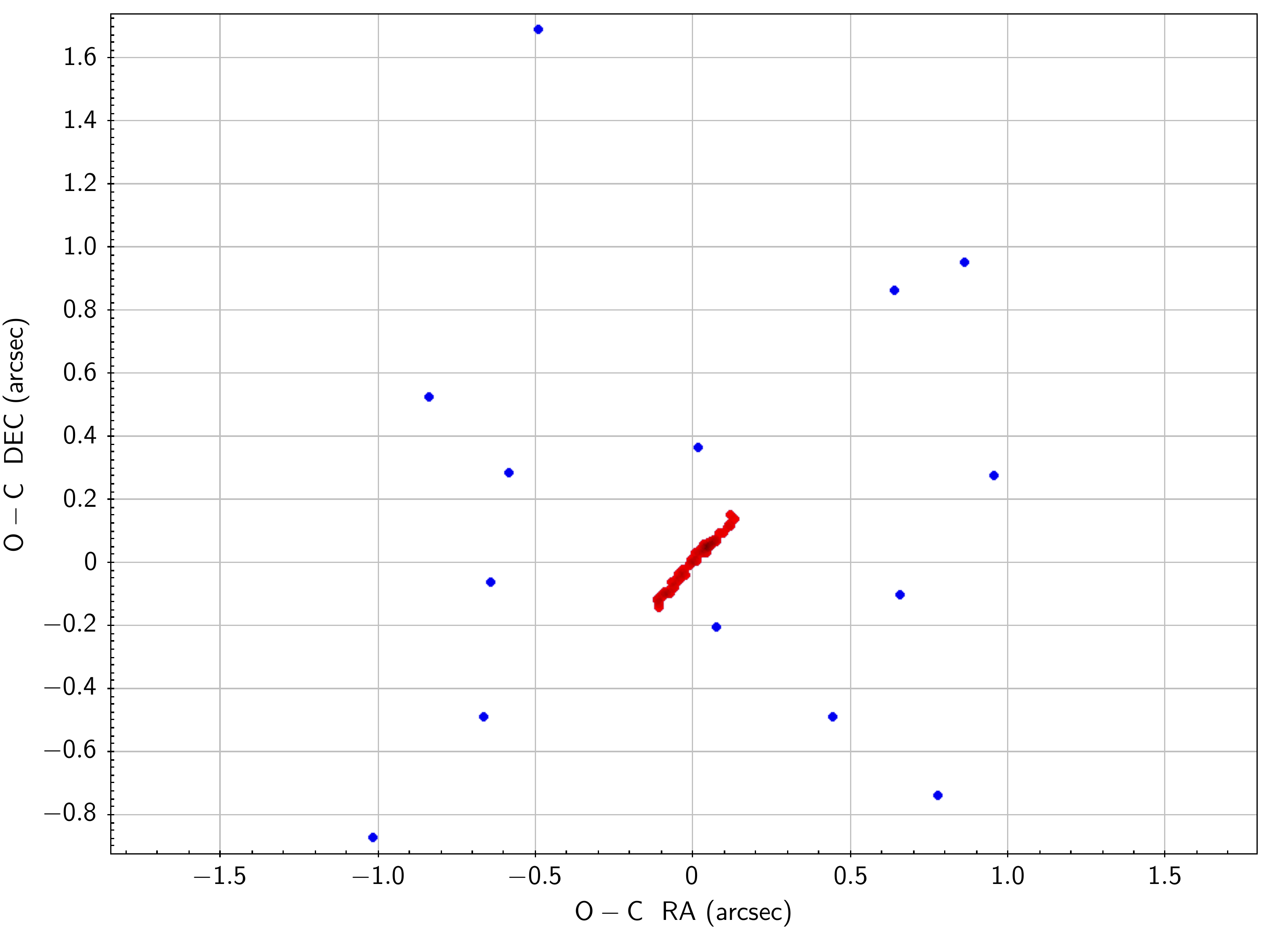}
    \caption{Residuals in right ascension and declination, in arcsec,
      obtained as a result of the orbit determination applied to the
      whole set of observations: Gaia (red points) and ground based
      (blue points) for the Gaia object $g1j0D7$.}
    \label{fig:g1j0D7_res}
  \end{figure}
\end{center}

The second Gaia object is $g1j0D7$, again a MBA. It has been observed by
Gaia on 2017, September 2. It has $7$ transits, which cover a time
span of $\sim 22$ hours. The object has follow-up observations from
the Abastuman Observatory (MPC code $119$) the night between September
10 and September 11. It obtained the MPC preliminary designation
$2017$~RW$_{16}$, but then it has been linked to the asteroid
$2006$~UL$_189$ discovered by the Catalina Sky Survey on June
  2005. Figure~\ref{fig:g1j0D7_AR} shows the result of the systematic
ranging when we use only Gaia observations. Again, we choose the point
with the minimum value of the $\chi^2$ as starting point, and we use
it as preliminary point. We then perform a differential corrections
least-squares fit, and we obtain the residuals (see
Fig.~\ref{fig:g1j0D7_res}), as described in the previous case.

\section{Future perspectives}

Gaia alerts runs daily, and they need a large effort to collect
follow-up observations from ground. With the method and the
  software described we have validated the already existing Java code
  written for the alert pipeline. Moreover, our approach can be also
  used to confirm the discoveries by computing an orbit using both
  Gaia and ground-based observations.

Then, when the accuracy for the short term will improve, we will also
be able to run the systematic ranging on less than three transits, given
the correct error model to the observations. Here it is worth noting
that the concept of short arc strongly depends on the concept of
curvature (see Sections~\ref{sec:cobweb} and \ref{sec:results}), which
is related not only to the time interval, but also to the accuracy of
the observations themselves.

The score computation represents a key feature in the Gaia frame (as
already pointed out in Sec.~\ref{sec:Gaia_ST}), but it will also be
very useful in future applications, like the ESA Euclid mission. Euclid
is an ESA mission with the aim of mapping the geometry of the dark
universe down to V$_{AB}$=24.5, with a launch scheduled for
  2020. While conducting its primary goal survey, Euclid will also
observe asteroids during his whole lifetime, and a Solar System
Working Group has been created within the Euclid consortium.

Euclid will observe at solar elongation $\sim 91^{\circ}$, and each
SSO will be imaged $16$ times (more precisely, $95\%$ of them $12$
times, and $50-60 \%$ of them $16$ times) over $67$
minutes~\citep{carry_euclid}. With a Hubble-like angular resolution,
astrometry will be quite good, at least at the same level of the best
observatories on ground (like Mauna Kea), or as the short term
accuracy of the Gaia alerts as it is now. The estimate accuracy is
around $100$ mas. While the southern sky will be repeatedly covered to
the same depth by LSST~\citep{lsst}, only Euclid will systematically
cover high declinations, with a strong potential for discoveries. For
each, having the score will be crucial to select the asteroid for
which trigger or not follow-up observations.

%------------------------%
%      CONCLUSIONS       %
%------------------------%
\section{Conclusions}
\label{sec:conclusions}

One of the main issue in the impact hazard assessment for imminent
impactors is given by the computation of the impact probability. The
main results of this article are a new algorithm to propagate the
probability density function from the space of the astrometric
residuals to the Manifold Of Variation, a geometric device to sample
the set of possible orbits, available even after a very short observed
arc.  In previous works, this computation was supported with an a
priori number density of asteroids. Our computation is complete,
rigorous, and uses no a priori hypothesis.

Does this new algorithm solve the problem of assessing the risk of
imminent impacts from a freshly discovered asteroid, with observations
limited to 1--2 tracklets? By using the AR and one of
our grid sampling, we have shown how to approximate a probability
integral on the portion of the MOV leading to an imminent impact, if
it is found. However, to accept this integral as Impact Probability we
need to check three conditions.

First, the probability density on the space of residuals needs to be
based upon a probabilistic model of the astrometric errors, taking
into account the past performances of the observatories. Second, the
observations used in the computation must be ``typical'' of the
observatory: even the best astronomical program produces a
comparatively small subset of ``faulty'' observations, with errors
much larger than the usual ones.  Third, we should assume that the
small sample of observations has statistical properties, such as mean
and standard deviation (STD), close to the ones of the full
distribution.

The first hypothesis is reasonable, in that a lot of work has been
developed in the last 20 years to produce astrometric error models for
asteroid observations (see \cite{carpino:outrej}, \cite{cbm10},
\cite{baer:err_mod}, \cite{farnocchia:fcct}). These models are not
perfect, but they represent an increasingly reliable source of
statistical information. The second hypothesis is not trivial: the
current format for asteroid observations does not contain sufficient
metadata to discriminate the ``weak observations'' from the good
ones. The full adoption of the new Astrometric Data Exchange Standard
(ADES\footnote{It is available at
  \url{http://minorplanetcenter.net/iau/info/IAU2015_ADES.pdf}.}),
approved by the IAU in 2015, will provide information such as SNR,
timing uncertainty, and so on, allowing to adapt the weighting of the
individual observations. The example of $P10vxCt$ shows how just one
lower quality observation can completely spoil the orbit results,
generating a false impact alarm. This can be avoided either with
remeasuring by the observer or by the orbit computer, provided such
down-weighting is supported by the metadata.

The third hypothesis is the most troublesome. Assuming that the
probability density of an astrometric error model is a perfect
statistical description, then by the law of large numbers a large
enough sample of $N$ observations shall have approximately the same
statistical properties of the model, with the differences going to
zero for $N\to +\infty$ (law of large numbers). Unfortunately,
$N=3,\ 4,\ 5$ is not large enough for the law of large numbers to
apply. For instance, a tracklet with $N=3$ observations can have all
the observations in one coordinate with errors $>2.5$ STD: this
statistical fluke would be very rare, occurring in a little more than
$1$ tracklet over $1$ million. Still, if a large asteroid survey
submits to the MPC more than one million tracklets per year, such a
fluke may occur about once a year, whereas the discovery of imminent
impactors is currently more rare (2 in 10 years). Detection of a rare
astronomical event cannot be a priori discriminated from rare
statistical events.

The tests on real cases discussed in this paper, and many more
from the NEOCP, convinced us that our algorithm
computes a reliable impact probability when the impact actually
occurs. Nevertheless, we cannot show that our algorithm is immune from
``false'' alarms. They are not false in the sense of a wrong
computation, or even worse a malicious disinformation, they are
statistical flukes which cannot be avoided because of lack of
information (hypothesis~2) and the need to use statistics on a small
sample (hypothesis~3). The question is what should be done to mitigate
the damage by these false alarms, given that we cannot avoid
disseminating them: otherwise, how could we disseminate the alarm in
the true case?

The only answer is to have a follow-up chain which does not waste
resources: the discoverers could themselves either remeasure or
follow-up on the short term, like 1 hour after discovery, the cases
announced as possible impactors.  Other telescopes should be available
to perform follow-up, to avoid improper use of survey telescopes for a
less demanding task. The ideal solution should be the availability of
a \textit{Wide Survey}, capable of covering the entire dark sky every
night and of detecting, e.g., an asteroid with absolute
magnitude $H=28$ at $0.03$ au distance (near opposition). Then the
same asteroid would be recovered by the survey the next day, before
the impact, and without the need for auxiliary follow-up.

\section*{acknowledgements}
This work has made use of data from the European Space Agency (ESA)
mission Gaia (\url{http://www.cosmos.esa.int/gaia}), processed by the
Gaia Data Processing and Analysis Consortium (DPAC,
\url{http://www.cosmos.esa.int/web/gaia/dpac/consortium}). Funding for
the DPAC has been provided by national institutions, in particular the
institutions participating in the Gaia Multilateral Agreement.

Part of this work is based on observations made at Observatoire de
Haute Provence (CNRS), France. The authors would also like to
acknowledge all the observers involved in the detection of the
asteroids discovered by Gaia: for g0T015 they are E.~Saquet,
M.~Dennefeld, D.~Gravallon, and for g1j0D7 they are R.~Y.~Inasaridze,
V.~Ayvazian, T.~Mdzinarishvili, Y.~Krugly.

F.~Spoto acknowledges support by the CNES post--doctoral program, and
A.~Del Vigna acknowledges support by the company SpaceDyS.
%\end{acknowledgements}

\appendix
%-------------------------%
%       APPENDIX: IP      %
%-------------------------%
\section{Probability density function computation}
\label{app:IPcomp}
In this appendix we give the mathematical details for the derivation
of equation \eqref{eq:p_S} for the probability density function on
the sampling space $S$.
\subsection{From the residuals space to the MOV}
\label{subsec:Res-MOV}
The first step of the procedure is the classical propagation of the
probability density function from the normalized residuals space to
the orbital elements space. From equation \eqref{eq:res_norm_gauss} we
recall that we start from the following density:
\[
  p_{\bm\Xi}(\bm \xi) = N(\bm 0,I_m)(\bm \xi)= \frac
  1{(2\pi)^{m/2}}\exp\left(-\frac 12 \bm\xi^T\bm\xi\right),
\]
We consider a point $\mathbf{x}_0\in \mathcal M$, the corresponding
image $\bm\xi_0 = F(\mathbf{x}_0)\in V$ and the tangent application
$DF(\mathbf{x}_0): T_{\mathbf{x}_0}\mathcal{X} \rightarrow
T_{\bm\xi_0}V$. Later we will discuss the choice of the point
$\mathbf{x}_0$. As proved in \citep[Section 5.7]{milani:orbdet}, we
obtain a probability density function on the orbital elements space
given by
\[
  p_{\bf {X}}(\mathbf{x}) = N(\mathbf{x}_0, \Gamma_{\mathbf X})(\mathbf{x}),
\]
where
$\Gamma_{\mathbf X} = P(\mathbf{x}_0)^{-1}(P(\mathbf{x}_0)^{-1})^T$ is
the covariance matrix resulting from the transformation of Gaussian
random variables under a linear transformation. Moreover, the normal
matrix of the variable $\mathbf{X}$ is
\[
  P(\mathbf{x}_0)^TP(\mathbf{x}_0) = B(\mathbf{x}_0)^TB(\mathbf{x}_0)=C(\mathbf{x}_0),
\]
namely the normal matrix of the constrained differential corrections
leading to $\mathbf{x}_0$, computed at convergence. Hence
\tiny
\begin{align*}
  p_{\mathbf{X}}(\mathbf{x}) &= \frac{\displaystyle \exp\left(-
                               \frac{mQ(\mathbf x)}{2}\right) |\det P(\mathbf{x}_0)|}
                               {\displaystyle\int_{T_{\mathbf{x}_0}\mathcal{X}} \exp \left(-
                               \frac{mQ(\mathbf y)}{2}\right) |\det
                             P(\mathbf{x}_0)|\,d\mathbf{y}}= \\
                           &= \frac{\displaystyle \exp \left(-
                               \frac{\chi^2(\mathbf x)}{2}\right)}
                               {\displaystyle\int_{T_{\mathbf{x}_0}\mathcal{X}}\exp \left(-
                               \frac{\chi^2(\mathbf y)}{2}\right) \,d\mathbf{y}}
\end{align*}
\normalsize
having used \eqref{chimov}, that is $m Q(\mathbf{x})= m
Q^*+\chi^2(\mathbf{x})$. Concerning the choice of $\mathbf{x}_0$ we
proceed as follows: if a reliable nominal solution $\mathbf{x}^\ast$
exists, we set $\mathbf{x}_0=\mathbf{x}^\ast$; if not, we select as
$\mathbf{x}_0$ the sample orbit having the minimum value of the
target function. This choice is also coherent with the $\chi$
computation given by \eqref{chimov}.

\subsection{From the MOV to the Admissible Region}
\label{subsec:MOV-AR}
We can now compute the determinant of the map $f_\mu$. Let $M_\mu$ be
the $2\times 2$ matrix representing the tangent map between $K'$ and
$\mathcal M$. $f_\mu$ is a
differentiable function, with Jacobian matrix
\[
B = Df_\mu = \frac{d\bold x}{d\bm\rho} =
\begin{pmatrix}
  \dfrac{\partial\mathcal A^*}{\partial\rho} & \dfrac{\partial\mathcal A^*}{\partial\dot\rho}\\[0.3cm]
  1 & 0\\
  0 & 1
\end{pmatrix}
=
\begin{pmatrix}
  \dfrac{\partial\mathcal A^*}{\partial\bm\rho}\\[0.3cm]
  I_2
\end{pmatrix}.
\]
We now consider $\bm\rho_0\in K'$ such that
$f_\mu(\bm\rho_0)=\mathbf{x}_0$. The matrix $B(\bm\rho_0)$ has rank 2,
thus $\mathcal M$ is smooth in the neighborhood of each of its
points. We can linearize in $\bm\rho_0$, obtaining the tangent map
\[
B(\bm\rho_0) = Df_\mu(\bm\rho_0) : T_{\bm\rho_0}K' \rightarrow T_{\mathbf{x}_0}\mathcal M,
\]
which is a linear map between two 2-dimensional spaces. We use a
rotation matrix $R$ in the orbital elements space, such that
\[
R(\bold x-\mathbf{x}_0) = \begin{pmatrix} \bold x'\\ \bold
  x''\end{pmatrix} \Rightarrow R^T \begin{pmatrix} \bold 0 \\ \bold
    x''\end{pmatrix} +\mathbf{x}_0 \in T_{\mathbf{x}_0}\mathcal M.
\]
It means that $\bold x''$ parametrizes $T_{\mathbf{x}_0}\mathcal
M$. In these coordinates the linearized map has a simple structure:
\[
RB(\bm\rho_0) = \begin{pmatrix} \bold 0\\ A(\bm\rho_0)\end{pmatrix},
\]
with $A(\bm\rho_0)$ an invertible $2\times 2$ matrix. By using that
$R$ is orthogonal, the following relation holds:
\begin{align*}
M_\mu^T M_\mu &= \left(RB(\bm\rho_0)\right)^T
\left(RB(\bm\rho_0)\right) = B(\bm\rho_0)^T (R^TR) B(\bm\rho_0) =\\ &=
B(\bm\rho_0)^T B(\bm\rho_0) = I_2+ \left(\dfrac{\partial\mathcal
  A^*}{\partial\bm\rho}(\bm\rho_0)\right)^T \dfrac{\partial\mathcal
  A^*}{\partial\bm\rho}(\bm\rho_0),
\end{align*}
and hence
\begin{equation}\label{eq:det_mmu}
\det M_\mu = \sqrt{\det\left(I_2+ \left(\dfrac{\partial\mathcal
  A^*}{\partial\bm\rho}(\bm\rho_0)\right)^T \dfrac{\partial\mathcal
  A^*}{\partial\bm\rho}(\bm\rho_0)\right)}.
\end{equation}

The next step is to explicitly compute the matrix
$\frac{\partial\mathcal A^*}{\partial\bm\rho}$. Hereinafter we neglect
terms containing the second derivatives of the residuals multiplied by
the residuals themselves. $\mathcal A^*(\bm\rho_0)$ is the
attributable which minimizes the target function $\left.Q(\mathcal
A,\bm \rho)\right|_{\bm\rho = \bm\rho_0}$. That is,
$\mathbf{x}_0=(\mathcal A^*, \bm\rho_0)$ is a zero
of the function
\[
F(\mathbf{x})=\frac{m}{2} \frac{\partial Q}{\partial\mathcal
  A}(\mathbf{x}) = B_\mathcal A(\mathbf{x})^T\bm\xi(\mathbf{x}).
\]
The function $F$ is continuously differentiable, and we have
\begin{align*}
  \frac{\partial F}{\partial \mathcal A}(\mathbf{x}) &=
  \frac{\partial}{\partial \mathcal A} \left(\frac{\partial \bm\xi}{\partial \mathcal A}(\mathbf{x})\right)^T \bm\xi(\mathbf{x}) + \left(\frac{\partial \bm\xi}{\partial \mathcal A}(\mathbf{x})\right)^T \frac{\partial\bm\xi}{\partial\mathcal A}(\mathbf{x})\simeq\\
  & \simeq \left(\frac{\partial \bm\xi}{\partial \mathcal A}(\mathbf{x})\right)^T \frac{\partial\bm\xi}{\partial\mathcal A}(\mathbf{x}) = C_{\mathcal A}(\mathbf{x}),
\end{align*}
where we used the approximation assumed at the beginning. The matrix
$C_{\mathcal A}(\mathbf{x}_0)$ is invertible, otherwise the doubly
constrained differential corrections would fail, and the minimum point
$\mathcal A^*$ could not be reached. By applying the implicit function
theorem, there exists a neighborhood $U$ of $\bm\rho_0$, a
neighborhood $W$ of $\mathcal A^*$, a continuously differentiable
function $\bold f:U\rightarrow W$ such that, for all $\bm\rho\in U$
holds
\[
 F(\mathcal A^*,\bm\rho)=\bold 0 \Leftrightarrow \mathcal A^*= \bold f(\bm\rho),
\]
and
\begin{equation}\label{eq:Dini}
\frac{\partial \bold f}{\partial \bm\rho}(\bm\rho) =
-\left(\frac{\partial F}{\partial \mathcal A}(\mathcal A^*(\bm\rho),\bm\rho)\right)^{-1}
\frac{\partial F}{\partial \bm\rho}(\mathcal A^*(\bm\rho),\bm\rho).
\end{equation}
We already computed $\frac{\partial F}{\partial \mathcal A}$, so we
proceed with the other derivative.
\begin{align*}
  \frac{\partial F}{\partial \bm\rho}(\mathbf{x}) &=
  \frac{\partial}{\partial \bm\rho} \left(\frac{\partial \bm\xi}{\partial \mathcal A}(\mathbf{x})\right)^T \bm\xi(\mathbf{x}) + \left(\frac{\partial \bm\xi}{\partial \mathcal A}(\mathbf{x})\right)^T \frac{\partial\bm\xi}{\partial\bm\rho}(\mathbf{x})\simeq\\
  & \simeq \left(\frac{\partial \bm\xi}{\partial \mathcal A}(\mathbf{x})\right)^T \frac{\partial\bm\xi}{\partial\bm\rho}(\mathbf{x}) = B_{\mathcal A}(\mathbf{x})^T B_{\bm\rho}(\mathbf{x}).
\end{align*}
By using the equation \eqref{eq:Dini} we obtain
\[
\frac{\partial \mathcal A^*}{\partial \bm\rho}(\bm\rho) = - C_{\mathcal
  A}(\mathcal A^*(\bm\rho),\bm\rho)^{-1}B_{\mathcal A}(\mathcal
A^*(\bm\rho),\bm\rho)^T B_{\bm\rho}(\mathcal A^*(\bm\rho),\bm\rho).
\]

\subsection{From the AR to the sampling space}
\label{subsec:AR-S}
The last step is the computation of the determinant of the map
$f_\sigma$, and this depends on $S$, for which we have different
possibilities. We call $M_\sigma$ the Jacobian matrix associated to
$f_\sigma$.

If the sampling is uniform in $\rho$, then $f_\sigma$ is the identity
map, and therefore $\det M_\sigma = 1$. If the sampling is uniform in
$\log_{10}(\rho)$, we have
\[
f_\sigma (\log_{10}(\rho),\dot\rho)=(\rho,\dot\rho),
\]
and hence
\[
M_\sigma = \begin{pmatrix} \log(10)\rho & 0\\0&1\end{pmatrix}
  \Rightarrow \det M_\sigma = \log(10)\rho.
\]
If we are in the cobweb case, the $f_\sigma$ is given by
\eqref{eq:cobweb}. Its Jacobian matrix is
\tiny
\[
M_\sigma =
\begin{pmatrix}
  \sqrt{\lambda_1}\cos\theta v_1^x - \sqrt{\lambda_2}\sin\theta v_1^y
& R\left[-\sqrt{\lambda_1}\sin\theta v_1^x - \sqrt{\lambda_2}\cos\theta v_1^y\right]\\
\sqrt{\lambda_2}\sin\theta v_1^x + \sqrt{\lambda_1}\cos\theta v_1^y
& R\left[\sqrt{\lambda_2}\cos\theta v_1^x - \sqrt{\lambda_1}\sin\theta v_1^y\right]
\end{pmatrix},
\]
\normalsize
where $\mathbf{v}_1 = (v_1^x,v_1^y)$. We have omitted the dependency
of $M_\sigma$ on $\bm\rho_0$ not to have a heavy notation. After some
manipulation, the determinant is
\[
\det M_\sigma = R\sqrt{\lambda_1\lambda_2} (\mathbf{v}_1\cdot \mathbf{v}_1).
\]

\bibliographystyle{aa}
\bibliography{immimp_biblio}

\begin{thebibliography}{32}
\expandafter\ifx\csname natexlab\endcsname\relax\def\natexlab#1{#1}\fi

\bibitem[{{Baer} {et~al.}(2011){Baer}, {Chesley}, \& {Milani}}]{baer:err_mod}
{Baer}, J., {Chesley}, S.~R., \& {Milani}, A. 2011, Icarus, 212, 438

\bibitem[{{Carpino} {et~al.}(2003){Carpino}, {Milani}, \&
  {Chesley}}]{carpino:outrej}
{Carpino}, M., {Milani}, A., \& {Chesley}, S.~R. 2003, Icarus, 166, 248

\bibitem[{{Carry}(2014)}]{carry2014_gfs}
{Carry}, B. 2014, in Gaia-FUN-SSO-3, 53

\bibitem[{{Carry}(2017)}]{carry_euclid}
{Carry}, B. 2017, Submitted to Astronomy \& Astrophysics

\bibitem[{{Chesley}(2005)}]{chesley2005}
{Chesley}, S.~R. 2005, in IAU Colloq. 197: Dynamics of Populations of Planetary
  Systems, ed. Z.~{Kne{\v z}evi{\'c}} \& A.~{Milani}, 255--258

\bibitem[{{Chesley} {et~al.}(2010){Chesley}, {Baer}, \& {Monet}}]{cbm10}
{Chesley}, S.~R., {Baer}, J., \& {Monet}, D.~G. 2010, Icarus, 210, 158

\bibitem[{Del~Vigna {et~al.}(2018)Del~Vigna, Milani, Spoto, Chessa, \&
  Valsecchi}]{delvigna2018}
Del~Vigna, A., Milani, A., Spoto, F., Chessa, A., \& Valsecchi, G.~B. 2018, in
  preparation

\bibitem[{{Farnocchia} {et~al.}(2015{\natexlab{a}}){Farnocchia}, {Chesley},
  {Chamberlin}, \& {Tholen}}]{farnocchia:fcct}
{Farnocchia}, D., {Chesley}, S.~R., {Chamberlin}, A.~B., \& {Tholen}, D.~J.
  2015{\natexlab{a}}, Icarus, 245, 94

\bibitem[{{Farnocchia} {et~al.}(2015{\natexlab{b}}){Farnocchia}, {Chesley}, \&
  {Micheli}}]{farnocchia2015}
{Farnocchia}, D., {Chesley}, S.~R., \& {Micheli}, M. 2015{\natexlab{b}},
  Icarus, 258, 18

\bibitem[{{Granvik} {et~al.}(2012){Granvik}, {Vaubaillon}, \&
  {Jedicke}}]{granvik2012}
{Granvik}, M., {Vaubaillon}, J., \& {Jedicke}, R. 2012, Icarus, 218, 262

\bibitem[{{Granvik} {et~al.}(2009){Granvik}, {Virtanen}, {Oszkiewicz}, \&
  {Muinonen}}]{granvik2009}
{Granvik}, M., {Virtanen}, J., {Oszkiewicz}, D., \& {Muinonen}, K. 2009,
  Meteoritics and Planetary Science, 44, 1853

\bibitem[{{Lindegren, L.} {et~al.}(2016){Lindegren, L.}, {Lammers, U.},
  {Bastian, U.}, {Hernández, J.}, {Klioner, S.}, {Hobbs, D.}, {Bombrun, A.},
  {Michalik, D.}, {Ramos-Lerate, M.}, {Butkevich, A.}, {Comoretto, G.},
  {Joliet, E.}, {Holl, B.}, {Hutton, A.}, {Parsons, P.}, {Steidelmüller, H.},
  {Abbas, U.}, {Altmann, M.}, {Andrei, A.}, {Anton, S.}, {Bach, N.}, {Barache,
  C.}, {Becciani, U.}, {Berthier, J.}, {Bianchi, L.}, {Biermann, M.},
  {Bouquillon, S.}, {Bourda, G.}, {Brüsemeister, T.}, {Bucciarelli, B.},
  {Busonero, D.}, {Carlucci, T.}, {Castañeda, J.}, {Charlot, P.}, {Clotet,
  M.}, {Crosta, M.}, {Davidson, M.}, {de Felice, F.}, {Drimmel, R.},
  {Fabricius, C.}, {Fienga, A.}, {Figueras, F.}, {Fraile, E.}, {Gai, M.},
  {Garralda, N.}, {Geyer, R.}, {González-Vidal, J. J.}, {Guerra, R.}, {Hambly,
  N. C.}, {Hauser, M.}, {Jordan, S.}, {Lattanzi, M. G.}, {Lenhardt, H.}, {Liao,
  S.}, {Löffler, W.}, {McMillan, P. J.}, {Mignard, F.}, {Mora, A.},
  {Morbidelli, R.}, {Portell, J.}, {Riva, A.}, {Sarasso, M.}, {Serraller, I.},
  {Siddiqui, H.}, {Smart, R.}, {Spagna, A.}, {Stampa, U.}, {Steele, I.},
  {Taris, F.}, {Torra, J.}, {van Reeven, W.}, {Vecchiato, A.}, {Zschocke, S.},
  {de Bruijne, J.}, {Gracia, G.}, {Raison, F.}, {Lister, T.}, {Marchant, J.},
  {Messineo, R.}, {Soffel, M.}, {Osorio, J.}, {de Torres, A.}, \& {O’Mullane,
  W.}}]{gaia_astrometry_2016}
{Lindegren, L.}, {Lammers, U.}, {Bastian, U.}, {et~al.} 2016, Astronomy and
  Astrophysics, 595, A4

\bibitem[{{LSST Science Collaborations} \& {LSST Project}(2009)}]{lsst}
{LSST Science Collaborations} \& {LSST Project}. 2009, {LSST Science Book}
  (Version 2.0, arXiv:0912.0201)

\bibitem[{{Mignard} {et~al.}(2007){Mignard}, {Cellino}, {Muinonen}, {Tanga},
  {Delb{\`o}}, {Dell'Oro}, {Granvik}, {Hestroffer}, {Mouret}, {Thuillot}, \&
  {Virtanen}}]{mignard2007}
{Mignard}, F., {Cellino}, A., {Muinonen}, K., {et~al.} 2007, Earth Moon and
  Planets, 101, 97

\bibitem[{{Milani} {et~al.}(2005{\natexlab{a}}){Milani}, {Chesley},
  {Sansaturio}, {Tommei}, \& {Valsecchi}}]{milani:clomon2}
{Milani}, A., {Chesley}, S.~R., {Sansaturio}, M.~E., {Tommei}, G., \&
  {Valsecchi}, G.~B. 2005{\natexlab{a}}, Icarus, 173, 362

\bibitem[{{Milani} \& {Gronchi}(2010)}]{milani:orbdet}
{Milani}, A. \& {Gronchi}, G.~F. 2010, {Theory of Orbit Determination}
  (Cambridge University Press)

\bibitem[{{Milani} {et~al.}(2004){Milani}, {Gronchi}, {De' Michieli Vitturi},
  \& {Kne\^zevi\'c}}]{milani2004AR}
{Milani}, A., {Gronchi}, G.~F., {De' Michieli Vitturi}, M., \& {Kne\^zevi\'c},
  Z. 2004, Celestial Mechanics and Dynamical Astronomy, 90, 57

\bibitem[{{Milani} {et~al.}(2008){Milani}, {Gronchi}, {Farnocchia}, {Kne{\v
  z}evi{\'c}}, {Jedicke}, {Denneau}, \& {Pierfederici}}]{milani2008}
{Milani}, A., {Gronchi}, G.~F., {Farnocchia}, D., {et~al.} 2008, Icarus, 195,
  474

\bibitem[{{Milani} {et~al.}(2007){Milani}, {Gronchi}, \& {Kne{\v
  z}evi{\'c}}}]{milani2007}
{Milani}, A., {Gronchi}, G.~F., \& {Kne{\v z}evi{\'c}}, Z. 2007, Earth Moon and
  Planets, 100, 83

\bibitem[{{Milani} \& {Kne{\v z}evi{\'c}}(2005)}]{milani2005}
{Milani}, A. \& {Kne{\v z}evi{\'c}}, Z. 2005, Celestial Mechanics and Dynamical
  Astronomy, 92, 1

\bibitem[{{Milani} {et~al.}(2005{\natexlab{b}}){Milani}, {Sansaturio},
  {Tommei}, {Arratia}, \& {Chesley}}]{milani2005multsol}
{Milani}, A., {Sansaturio}, M.~E., {Tommei}, G., {Arratia}, O., \& {Chesley},
  S.~R. 2005{\natexlab{b}}, Astronomy \& Astrophysics, 431, 729

\bibitem[{{Milani} \& {Valsecchi}(1999)}]{milani99}
{Milani}, A. \& {Valsecchi}, G.~B. 1999, Icarus, 140, 408

\bibitem[{{Muinonen} \& {Bowell}(1993)}]{muinonen93}
{Muinonen}, K. \& {Bowell}, E. 1993, Icarus, 104, 255

\bibitem[{{Muinonen} {et~al.}(2016){Muinonen}, {Fedorets}, {Pentik{\"a}inen},
  {Pieniluoma}, {Oszkiewicz}, {Granvik}, {Virtanen}, {Tanga}, {Mignard},
  {Berthier}, {Dell`Oro}, {Carry}, \& {Thuillot}}]{muinonen2016}
{Muinonen}, K., {Fedorets}, G., {Pentik{\"a}inen}, H., {et~al.} 2016, Planetary
  and Space Science, 123, 95

\bibitem[{{Muinonen} {et~al.}(2001){Muinonen}, {Virtanen}, \&
  {Bowell}}]{muinonen2001}
{Muinonen}, K., {Virtanen}, J., \& {Bowell}, E. 2001, Celestial Mechanics and
  Dynamical Astronomy, 81, 93

\bibitem[{{Oszkiewicz} {et~al.}(2009){Oszkiewicz}, {Muinonen}, {Virtanen}, \&
  {Granvik}}]{oszkiewicz2009}
{Oszkiewicz}, D., {Muinonen}, K., {Virtanen}, J., \& {Granvik}, M. 2009,
  Meteoritics and Planetary Science, 44, 1897

\bibitem[{{Prusti}(2012)}]{prusti2012}
{Prusti}, T. 2012, Astronomische Nachrichten, 333, 453

\bibitem[{{Tanga} {et~al.}(2007){Tanga}, {Delb{\`o}}, {Hestroffer}, {Cellino},
  \& {Mignard}}]{tanga2007}
{Tanga}, P., {Delb{\`o}}, M., {Hestroffer}, D., {Cellino}, A., \& {Mignard}, F.
  2007, Advances in Space Research, 40, 209

\bibitem[{{Tanga} {et~al.}(2016){Tanga}, {Mignard}, {Dell`Oro}, {Muinonen},
  {Pauwels}, {Thuillot}, {Berthier}, {Cellino}, {Hestroffer}, {Petit}, {Carry},
  {David}, {Delbo`}, {Fedorets}, {Galluccio}, {Granvik}, {Ordenovic}, \&
  {Pentik{\"a}inen}}]{tanga2016}
{Tanga}, P., {Mignard}, F., {Dell`Oro}, A., {et~al.} 2016, Palnetary and Space
  Science, 123, 87

\bibitem[{{Thuillot} {et~al.}(2014){Thuillot}, {Carry}, {Berthier}, {David},
  {Hestroffer}, \& {Rocher}}]{thuillot2016}
{Thuillot}, W., {Carry}, B., {Berthier}, J., {et~al.} 2014, in SF2A-2014:
  Proceedings of the Annual meeting of the French Society of Astronomy and
  Astrophysics, ed. J.~{Ballet}, F.~{Martins}, F.~{Bournaud}, R.~{Monier}, \&
  C.~{Reyl{\'e}}, 445--448

\bibitem[{Tommei(2006)}]{tommei:phd}
Tommei, G. 2006, PhD thesis, University of Pisa

\bibitem[{{Virtanen} {et~al.}(2001){Virtanen}, {Muinonen}, \&
  {Bowell}}]{virtanen2001}
{Virtanen}, J., {Muinonen}, K., \& {Bowell}, E. 2001, Icarus, 154, 412

\end{thebibliography}
\end{document}